\documentclass[%
 reprint,superscriptaddress,
 amsmath,amssymb,nofootinbib,
 aps,twocolumn,pra
]{revtex4-2}
\usepackage[utf8]{inputenc}
\usepackage{bbold}

\newcommand{\eqn}[1]{\begin{eqnarray} #1 \end{eqnarray}}
\newcommand{\tit}[1]{\textit{#1}}
\newcommand{\tbf}[1]{\textbf{#1}}
\newcommand{\trm}[1]{\textrm{#1}}

\newcommand{\tr}[1]{  \textrm{tr}\left[ #1 \right]  }
\newcommand{\trc}[2]{  \textrm{tr}_{#1}\left[ #2 \right]  }
\newcommand{\zum}[2]{\displaystyle\sum_{#1}^{#2}}

\newcommand{\ket}[1]{| #1 \rangle}
\newcommand{\ketbra}[2]{| #1 \rangle \langle #2 |}
\newcommand{\braket}[2]{\langle #1 |  #2 \rangle}

\newcounter{theorem}
\newenvironment{theorem}[1][]{\refstepcounter{theorem}\par\medskip
   \noindent\textbf{Theorem~\thetheorem. #1} \rmfamily}{\medskip}
\newcommand{\thrm}[1]{\begin{theorem} #1 \end{theorem}}

\newcounter{proposition}
\newenvironment{proposition}[1][]{\refstepcounter{proposition}\par\medskip
   \noindent\textit{Proposition~\theproposition #1:} \rmfamily}{\medskip}
\newcommand{\prop}[1]{\begin{proposition} #1 \end{proposition}}

\newcounter{definition}
\newenvironment{definition}[1][]{\refstepcounter{definition}\par\medskip
   \noindent\textbf{Definition~\thedefinition. #1} \rmfamily}{\medskip}
\newcommand{\defn}[1]{\begin{definition} #1 \end{definition}}

\begin{document}

\title{Quantum dynamics is linear because quantum states are epistemic}

\author{Jacques L. Pienaar\smallskip}
\affiliation{
\href{http://www.physics.umb.edu/Research/QBism/}{QBism Group, University of Massachusetts Boston}, 100 Morrissey Boulevard, Boston MA 02125, USA}

\begin{abstract}
  According to quantum theory, a scientist in a sealed laboratory cannot tell whether they are inside a superposition or not. Consequently, so long as they remain isolated, they can assume without inconsistency that their measurements result in definite outcomes. We elevate this to the status of a general principle, which we call Local Definiteness. We apply this principle in the context of modifications of quantum theory that allow the dynamics to be non-linear. We prove that any such theory satisfies Local Definiteness if and only if its dynamics is linear. We further note that any interpretation that takes quantum states to be epistemic necessarily satisfies the principle, whereas interpretations that take quantum states to be ontic do not satisfy it, unless they make additional assumptions that amount to presupposing linearity of the dynamics. Therefore the reason why experiments to date have not found evidence of non-linear dynamics might simply be that quantum states are epistemic.  
\end{abstract}

\maketitle

\section{Introduction}

Quantum theory involves two notions of \tit{linearity}: a kinematical one and a dynamical one. \tit{Kinematical linearity} refers to the fact that the pure states of the theory can be represented (up to an arbitrary choice of phase) by a vector $\ket{\Psi }$ in a Hilbert space, such that for any set of valid physical states $\{ \ket{\Psi_j } \}_j$ one can construct another valid state by making a \tit{coherent superposition}:
\eqn{ \label{eqn:linear-kinematics}
 \ket{\Phi} := \zum{j}{} \, \alpha_j \, \ket{\Psi_j} 
}
where the complex coefficients $\alpha_j$ may be chosen arbitrarily up to the constraint that the resulting state be normalized\footnote{In the special case where the $\ket{\Psi_j }$ are orthonormal, this reduces to the requirement that $\sum_j |\alpha_j|^2 = 1$.}, $|\braket{\Phi}{\Phi}|^2=1$. By comparison, what we shall call \tit{dynamical linearity} consists in the fact that the evolution of an isolated pure state $\ket{\Psi_j }$ over some time interval $T$ is represented by a unitary map $U_{T}$ which acts \tit{linearly} on the states:
\eqn{ \label{eqn:linear-dynamics}
\ket{\Phi(T)} &:=& U_{T} \ket{ \Phi(0) } \nonumber \\
&=& U_{T} \left( \zum{j}{} \, \alpha_j \, \ket{\Psi_j (0)} \right) \nonumber \\
&=& \zum{j}{} \, \alpha_j \, U_{T} \ket{\Psi_j(0)}  \nonumber \\
&:=&  \zum{j}{} \, \alpha_j \, \ket{\Psi_j(T)} \, .
}
Consequently, if we are given a set of solutions $\{ \Psi_j(T) \}_j$ to the dynamical equations as represented by the map $U_{T}$, we can always construct another solution $\ket{\Phi(T)}$ to the same dynamics by superposing the known solutions as per the final line of Eq.~\eqref{eqn:linear-dynamics} (this is what textbooks usually call ``the superposition principle"). 

These two notions are conceptually distinct. Kinematical linearity is essentially a constraint on the set of valid physical states: it says that if certain vectors represent valid physical states, then there must be a valid physical state represented by the superposition of these vectors. By contrast, dynamical linearity represents a constraint on the dynamics of the theory.  

Indeed, the two notions are independent: either one can be false while the other remains true. For so long as we accept the conceptual distinction between kinematics and dynamics, it is at least logically possible that future experiments might reveal deviations from quantum \tit{dynamics} -- such as small non-linear corrections to the dynamical equations -- without however affecting the \tit{kinematics} of the theory. More specifically, we can consistently suppose that all coherent superpositions of physical states yield physical states, while at the same time allowing the dynamics to violate Eq.~\eqref{eqn:linear-dynamics}, so that taking a superposition over the inputs to a dynamical process is not equivalent to taking the same superposition over the outputs of the process. 

This observation has inspired numerous theoretical models describing possible deviations from dynamical linearity\footnote{See for example Refs~\cite{ Bassi_review,collapse_SEP,Hejlesen_thesis,Pienaar_thesis,Kent_2021} and citations therein.}. It is noteworthy that experiments to test these proposals have so far not found any evidence of such effects. This is striking, considering that all previous paradigmatic examples of physical theories -- such as fluid mechanics, electrodynamics and gravity -- are governed by non-linear dynamical equations. Why should quantum theory be different?

The linearity of quantum dynamics has an interesting consequence: it ensures that a scientist enclosed within an isolated laboratory cannot tell by any local experiment (i.e.\ performed within the laboratory) whether or not the whole laboratory and its contents including the scientist themselves are `inside a superposition'. Though rarely remarked upon, this curious fact dates back as far as Hugh Everett III's 1957 article on the `relative state interpretation', in which he argued that, due to the linearity of the quantum wave equation, an observer who enters a superposition state should not ``feel themselves split"~\cite{Everett57,Barrett_SEP}. 

One can also establish the same result by proving the contra-positive: if it were possible for the encapsulated observer to determine by some local process whether or not the entire laboratory including themselves had entered a superposition state, this implies they would have the ability to perfectly distinguish between the total superposition state and a state representing just one of its components. It then suffices to note that since these states are non-orthogonal, they cannot be perfectly distinguished by any linear process.

In this paper we aim to establish the implication in the opposite direction: we aim to prove that \tit{if} the laws of physics forbid us from determining by local experiments that we are inside a superposition, then the dynamical laws must be linear. We will then argue that this result has consequences for the interpretation of quantum states.

The logical structure of our argument runs as follows: we consider the following empirical principle that we call \tit{Local Definiteness} (LD): \tit{a scientist confined to an enclosed laboratory cannot determine by any local experiment whether the laboratory and all its contents are in a definite effectively classical state, or in a superposition of such states}. We then ask why it should hold true. The standard justification, as we have seen, assumes that quantum dynamics is linear and derives LD as a consequence. The problem with this strategy is that it leaves no room for contemplating more general theories in which the dynamics might not be linear.

We shall therefore introduce a more general framework of probabilistic theories in which the dynamics need not be linear, for which LD may not necessarily hold. Our main result is to prove that, within this more general framework, a theory satisfies LD only if its dynamics is linear. 

This result has implications for the foundational question of whether quantum states are epistemic (representing knowledge or beliefs) or ontic (representing states of affairs in reality). As we will see, epistemic interpretations of quantum states imply that LD holds; conversely ontic interpretations are not so constrained and can allow non-linear dynamics. Our theorem therefore implies that dynamics is necessarily linear if one regards quantum states as epistemic. We argue that this makes epistemic interpretations superior to ontic interpretations at explaining why quantum dynamics is linear. 


The paper is structured as follows. In Sec~\ref{sec:warmup} we prove that if we assume only the kinematical structure of quantum theory, then the principle of Local Definiteness is enough to imply the linearity of quantum dynamics. In Sec~\ref{sec:main} we show that a stronger result holds: given a probabilistic theory whose kinematics is not necessarily equivalent to that of quantum theory, Local Definiteness implies that the dynamics must act linearly on convex mixtures of probabilistic states. In Sec~\ref{sec:epistemic} we argue that, by application of Leibniz' principle, the absence of experimental evidence for non-linear quantum dynamics is best explained by adopting an epistemic interpretation of quantum states.

\section{A primer: quantum theory with non-linear dynamics \label{sec:warmup}}

As a warm-up exercise, let us assume that the kinematics of our theory is equivalent to that of quantum theory: we associate a unique density operator $\rho$ to every physical state, and a POVM $M := \{M^{(k)} \}_k$ to every measurement, where $k \in \{1,2,\dots \}$ runs over the outcomes and $\sum_k \, M^{(k)} = \mathbb{1}$. Pure states are rank-1 projectors $\rho = \ketbra{\psi}{\psi}$ and the kinematical superposition principle holds, ie.\ if $\{ \ket{\Psi_j } \}_j$ represent valid states (up to an overall phase), then any superposition of them is also a valid state, as per \eqref{eqn:linear-kinematics}. Probabilities are calculated using the Born rule,
\eqn{ 
\trm{Pr}(k|\rho, M) = \tr{\rho M^{(k)}} \, .
}
The only difference to standard quantum theory is that we shall permit dynamical transformations of states to be non-linear in general:

\defn{\label{defn:convex-linear-density}Consider a transformation $T(\rho)$ that maps density matrices to density matrices. We say that $T$ is \tit{convex-linear on density matrices} iff  
\eqn{ \label{eqn:convex-linear-density}
T\left( \zum{k}{} \, \lambda_k \, \rho_{k} \right) = \zum{k}{} \, \lambda_k \, T(\rho_{k}) \, 
}
for arbitrary convex coefficients $\lambda_k > 0$, $\sum_k \, \lambda_k = 1$. Otherwise we say $T$ is \tit{non-convex-linear}. }

Consider now a system $\mathcal{S}_1$ sealed inside a laboratory with a scientist, modeled as another system $\mathcal{S}_2$. For simplicity we assume that all other parts and contents of the lab can be either ignored or subsumed into one of these two systems. Suppose the scientist $\mathcal{S}_{2}$ performs operations that -- if we were to perform them ourselves -- we would regard as constituting a measurement of a POVM $M_1 := \{ M^{(k)}_{1} \}_k$ on system $\mathcal{S}_{1}$. A question arises: given that it is not us but $\mathcal{S}_2$ performing these operations on $\mathcal{S}_1$, what state should we assign the total system from our standpoint outside the lab?

This is precisely the scenario considered by Wigner in his famous thought experiment, and the reason it has caused such controversy is that the answer hinges on the contentious issue of precisely when a measurement should be judged to have occurred. Does $\mathcal{S}_{2}$'s activity inside the isolated lab also collapse the state for us on the outside? Or should we model it as a coherent entangled superposition? We will not be concerned here with trying to settle this debate, but shall focus on a much simpler question: can the scientist inside the lab tell the difference between the two options, without help from the outside?

First note that if their measurement has indeed `collapsed' the state, then we on the outside may reasonably assign the total system $\mathcal{S}_{1,2}$ (representing the lab and its contents) a state of the form $\ket{k}_1 \ket{\phi^{(k)}}_2$, where $\ket{\phi^{(k)}}_2$ may be interpreted as the state assigned to $\mathcal{S}_{2}$ just in case we are sure that they would report having observed outcome $k$, if interrogated\footnote{All that matters for our purposes is that we have some way of checking what $\mathcal{S}_{2}$ observed. This might entail just asking $\mathcal{S}_{2}$ what they observed and interpreting their response; or it might require more complicated actions involving truth serum or neurosurgery; these details are unimportant to our argument.}. But since we on the outside do not know which outcome $k$ was actually obtained, we must assign some probabilities $\{ \lambda_k \}_k$ representing our uncertainty about the value of the outcome. Accordingly, we assign not a single state to the system but an \tit{epistemic ensemble} of states, represented as\footnote{A more thorough treatment of ensemble notation will be given in Sec~\ref{sec:main}.}
\eqn{ \label{eqn:total_ensemble_baby}
E_{1,2} &:=&  \{ \lambda_k : \ketbra{k}{k}_{1} \otimes \ketbra{\phi^{(k)}}{\phi^{(k)}}_{2} \}_k \nonumber \\
&:=& \{ \lambda_k : \rho^{(k)}_{1,2} \}_k \, .
}
By definition the statistics of such an epistemic ensemble for an arbitrary POVM $M = \{ M^{(j)} \}_j$ are calculated as:
\eqn{
 \trm{Pr}(j|E_{1,2},M) &:=& \zum{k}{} \, \lambda_k \, \trm{Pr}(j|\rho^{(k)}_{1,2},M) \nonumber \\
 &=& \zum{k}{} \, \lambda_k \, \tr{\rho^{(k)}_{1,2} M^{(j)} } \, . 
}
Note that in standard quantum theory, an arbitrary ensemble $E := \{ \lambda_k : \rho^{(k)}\}_k$ can be represented by a unique density operator $\varrho(E)$ via the mapping
\eqn{ \label{eqn:varrho}
\varrho(E) :=  \zum{k}{} \, \lambda_k \,  \rho^{(k)} \, ,
}
since the latter evidently produces the same statistics for any measurement (due to the linearity of the trace). For this reason, mixed density operators are often referred to as `ensembles'. 

However, since we shall be going beyond standard quantum theory by contemplating non-convex-linear dynamics $T$, we must be careful to make a formal distinction between purely epistemic combinations of states (also known as `proper mixtures'), versus `improper mixed states', which represent uncertainty that is (possibly) attributable to the system itself independently of what anybody knows about it\footnote{The conceptual distinction between proper and improper mixtures was emphasized by d’Espagnat~\cite{dEspagnat}; see also Timpson \& Brown~\cite{Timpson_Brown_2004}.}. 

The reason why we need to make this formal distinction becomes clear when we consider how a non-linear transformation acts upon a given state. On one hand, intuitively $T$ should act on an ensemble $E = \{ \lambda_k : \rho^{(k)} \}_k$ by acting individually on each of its components\footnote{See the discussion of `quasi-linear dynamics' in Appendix \ref{app:quasilinear} for more explanation of why we regard this rule as `intuitive'.}:
\eqn{ 
T(E) :=  \{ \lambda_k : T(\rho^{(k)}) \}_k \, .
}
If on the other hand we were to associate the ensemble $E$ with a density matrix $\varrho(E)$ via the usual map \eqref{eqn:varrho}, we would find that $T(\varrho(E)) \neq \varrho(T(E))$ in general. In other words, if we try to compute the time evolution of $E$ by computing $T(\varrho(E))$, we do not get the correct answer, namely, we do not obtain a density matrix which is statistically equivalent to the time-evolved ensemble $T(E)$. To be precise: 

\prop{\label{prop:linear-condition} The map $T$ satisfies $T(\varrho(E)) = \varrho(T(E))$ for all ensembles $E$ iff $T$ is convex-linear on density operators per Definition~\ref{defn:convex-linear-density}. (Proof: just let $E = \{ \lambda_k : \rho_k \}_k$ and observe that the LHS of Eq~\eqref{eqn:convex-linear-density} is $T(\varrho(E))$ and the RHS is $\varrho(T(E))$).
}

Therefore, to avoid ambiguity, throughout this article we use the term `ensemble' exclusively for epistemic ensembles as in Eq~\eqref{eqn:total_ensemble_baby}, and we use density operators exclusively for improper mixed states, whose interpretation is left unspecified.

Turning now to the second option, if we think that $\mathcal{S}_{2}$'s measurement has not really collapsed the state but has only caused $\mathcal{S}_{2}$ to become entangled to system $\mathcal{S}_{1}$, it would be reasonable to assign the total system a pure state of the form $\rho^{(\trm{+})}_{1,2} = \ketbra{\Psi^{(+)}}{\Psi^{(+)}}_{1,2}$ where:
\eqn{ \label{eqn:superposition_baby}
 \ket{\Psi^{(+)}}_{1,2} &:=& \left( \zum{k}{} \, \alpha_k \ket{k}_{1}\ket{\phi^{(k)}}_{2} \right) \, ,
}
and where the complex amplitudes yield the expected probabilities, $|\alpha_k|^2 = \lambda_k$. 

The ensemble $E_{1,2}$ and superposition $\rho^{(\trm{+})}_{1,2}$ imply different statistics in general, so we can in principle differentiate them by doing appropriate measurements on the total system. But can the scientist $\mathcal{S}_{2}$ inside the lab distinguish them? 
To answer this, we shall make the following assumption:\\

\tbf{Assumption:} Any and all operations carried out by $\mathcal{S}_{2}$ inside the laboratory can be modeled by us on the outside as some combination of a transformation followed by a measurement on system $\mathcal{S}_{1}$ .
\\

We justify this assumption as follows. Intuitively, the fact that $\mathcal{S}_{2}$ is `sealed inside a laboratory' means that the set of operations they can carry out is limited, compared to what we could do (in principle) from our standpoint outside the lab\footnote{As a trivial example, any procedure that would involve immobilizing or otherwise totally incapacitating $\mathcal{S}_{2}$ for the duration is obviously beyond their own ability to perform. The best they could do is delegate the task to an external person or mechanism, which is precisely the role played by `us' outside the laboratory.}. In order to represent this limitation formally, we propose that the operations achievable by $\mathcal{S}_{2}$ inside the lab must be modeled as a strict subset of the operations that we can perform from the outside. We can then conventionally \tit{define} the subsystem $\mathcal{S}_{1}$ as comprising those degrees of freedom which are targeted by this subset of operations, and $\mathcal{S}_{2}$ as comprising the (assumed non-trivial) remainder. 

We emphasize that this is an extremely weak assumption, not to be confused with a significantly stronger assumption sometimes made in the literature, which says that a scientist can only apply a physical theory to systems external to themselves. Indeed, at no point in our argument do we attempt to adopt the point of view of $\mathcal{S}_{2}$, nor do we make any assumptions about how \tit{they} should model what happens to them. In particular our analysis is consistent with proposals where $\mathcal{S}_{2}$ assigns a superposition state to themselves in order to make predictions, such as Ref~\cite{Baumann2020}.

Continuing with the thought experiment, let us therefore suppose there exist some operations achievable by $\mathcal{S}_{2}$ that would reveal to us whether the total state is the ensemble $E_{1,2}$ or the superposition $\rho^{(\trm{+})}_{1,2}$. By the above assumption, from the outside we should be able to model these operations as a special transformation $\tilde{T}$ on $\mathcal{S}_{1}$ followed by a special POVM $\tilde{M} = \{ \tilde{M}^{(j)} \}_j$, such that the ensemble and superposition states imply different probabilities for the outcomes of $j$ (for if the statistics did not differ between the two cases, then $\tilde{M}$ would be powerless to reveal any difference between them).

First consider the special case where $\tilde{T}$ is trivial, and ask whether one could succeed at the task just by performing some measurement $\tilde{M}$ on $\mathcal{S}_{1}$. It is easy to see that this will not work, because the reduced density matrix associated to $\mathcal{S}_{1}$ is the same in either case. For if the total state is $\rho^{(\trm{+})}_{1,2}$ then the statistics are determined from the reduced density matrix: 
\eqn{ \label{eqn:T-state-baby}
\rho_1 &:=&  \trc{2}{ \rho^{(\trm{+})}_{1,2} } \nonumber \\
&=& \zum{k}{} \, \lambda_k \, \ketbra{k}{k}_1 \, .
}
On the other hand if the total system is the ensemble $\tilde{E}_{1,2}$ then system $\mathcal{S}_{1}$ alone is assigned the ensemble $E_1 := \{ \lambda_k : \ketbra{k}{k}_1 \}_k$, and since $\varrho(E_1) = \rho_1$, this yields the same statistics for all measurements.

The above argument implies that any protocol which would allow $\mathcal{S}_{2}$ to succeed at their task must involve a non-trivial transformation $\tilde{T}$. Moreover, success requires that $\varrho(\tilde{T}(E_1)) \neq \tilde{T}(\rho_1)$, since otherwise the reduced states of $\mathcal{S}_{2}$ would still remain indistinguishable. In fact, since $\rho_1 = \varrho(E_1)$, the requirement can be written as $\varrho(\tilde{T}(E_1)) \neq \tilde{T}(\varrho(E_1))$; it then follows from Proposition~\ref{prop:linear-condition} that $\tilde{T}$ must be non-linear.

We conclude that it is impossible for the scientist in the closed laboratory to determine whether or not they are inside a superposition (that is, Local Definiteness holds) iff the allowed transformations $T$ are convex-linear on density matrices\footnote{On the face of it, this might appear to contradict another result in the literature: the demonstration of `quasi-linear' transformations, which preserve the convex structure of the space of density matrices but are not linear~\cite{Rembielinski_Caban_2020,Rembielinski_Caban_2021}. We explain in Appendix~\ref{app:quasilinear} how such transformations are compatible with our present results.}. 

For completeness, we would like to show that our general definition of convex-linearity encompasses unitary dynamics for pure states. Fortunately this work has already been done: in 2009 T.F. Jordan pointed out that if we assume\footnote{Jordan actually did not think any extra assumption was needed, for he assumed that invertibility of the map followed from its being linear and injective. The error was later pointed out by Hejlesen~\cite{Hejlesen_thesis}.} the convex-linear map $T$ is invertible, then it must transform pure states into pure states and preserve their Hilbert-Schmidt inner products. One can then use Bargmann's refinement of Wigner's theorem~\cite{Bargmann,WignerBook} to prove that $T$ must be either unitary or anti-unitary, and in particular it must have the form
 \eqn{
 T( \rho ) = U \rho U^{\dagger}
 }
 where $U$ acts linearly on the pure states as per \eqref{eqn:linear-dynamics}.

Straightforward as it is, the argument of this section is still too weak: a critic could argue that by assuming the kinematical structure of quantum theory we are `sneaking linearity in by the back door'. After all, maybe the linearity of quantum state space alone is doing all the heavy lifting in the foregoing argument. To close this loophole, in the next section we extend the argument to a general class of probabilistic theories, making minimal assumptions about their kinematics and dynamics. This will allow us to subsequently establish a firm link between the principle of Local Definiteness and dynamical linearity.

\section{General version of the argument \label{sec:main}}

\subsection{Preliminary definitions \label{sec:framework}}

\subsubsection{Operational theories}
To go beyond quantum theory while making the fewest assumptions possible, we appeal to the concept of an \tit{operational theory}, first introduced in Refs.\ \cite{Spekkens05,Harrigan_Spekkens}. Briefly, the central object of an operational theory is an abstract \tit{experiment} consisting in a sequence of operations, specifically a \tit{preparation} $P$, a \tit{transformation} $T$, and a \tit{measurement} $M$, culminating in the realization of some \tit{outcome} $k$ from a discrete finite set of possible values $\{1,2,\dots K\}$. The operations $P,T,M$ are to be thought of as descriptions of certain procedures to be carried out by the scientist in a suitably equipped laboratory. For every experiment on a given system, defined by some triple $(P,T,M)$, the operational theory assigns probabilities $\trm{Pr}(k|P,T,M)$.
In general, a list of laboratory procedures contains much more information than is relevant to the statistical predictions of the operational theory. For example, given some quantum state $\ket{\psi}$, there will in general be many ways to prepare this same state, depending on the particular laboratory implementation. We are therefore motivated to define a \tit{state} $S$ in the operational theory as an \tit{equivalence class} of preparation procedures, all of which give identical predictions in the operational theory. Formally, if two preparation procedures $P,P'$ produce the same probabilities for all combinations of transformations and measurements, \tit{ie}.\ if 
\eqn{
\trm{Pr}(k|P,T,M) = \trm{Pr}(k|P',T,M) \qquad \forall T,M \, ,
}
then we shall use a single symbol $S$ to represent both procedures, referring to $S$ as the state assigned to the system. By this language we do not mean to imply that it represents some \tit{state of affairs} in reality, but nor do we exclude such an interpretation. 

Indeed, throughout this article we shall remain noncommittal about the precise meaning of the symbols $S,T,M$. One could interpret them as \tit{ontic states} that describe or represent states of affairs in reality; or one could treat them \tit{instrumentally} as mere symbols that allow us to make predictions via the operational theory, and nothing more. None of the results we shall prove in this article will depend on which interpretation one adopts. 

It is common in the literature to augment operational theories with an additional \tit{compositional} structure that tells us how to compose smaller experiments into larger ones and vice-versa. We shall not need to appeal to any particular compositional framework; however we shall require at least a concept of \tit{parallel composition} of processes, which we now describe. Given any experiment $(S_1,T_1,M_1)$ resulting in outcome $k_1$, and $(S_2,T_2,M_2)$ resulting in outcome $k_2$, defined on subsystems 1 and 2 respectively, the procedure in which both procedures are carried out in parallel is represented by $(S_1,S_2,T_1,T_2,M_1,M_2)$, and the joint probabilities for outcomes $k_1$, $k_2$ are given by the product:
\eqn{ \label{eqn:factorize-P}
&& \trm{Pr}(k_1 k_2|S_1,S_2,T_1,T_2,M_1,M_2) \nonumber \\
&&:= \trm{Pr}(k_1|S_1,T_1,M_1) \, \trm{Pr}(k_2|S_2,T_2,M_2) \, .
}
Finally, it is useful to separate kinematics from dynamics, using the convention that one can simply drop the `$T$' whenever the transformation is trivial; thus the pair $(S,M)$ indicates that a system assigned state $S$ is immediately measured by $M$ without being transformed (say, by evolving over time) in between. Statements referring to such state-measurement pairs can then be thought of as capturing the \tit{kinematical} content of the theory, whereas statements concerning the possible non-trivial transformations $T$ constitute the \tit{dynamical} content of the theory.

\subsubsection{Epistemic layer}
An important innovation we shall make, which goes beyond previous literature, is to introduce what we call an \tit{epistemic layer} on top of the operational theory. Intuitively, as we have said, the operational theory is ambiguous as to whether the states $S$ have an ontological significance. Nevertheless, there are always situations in which our uncertainty is \tit{unambiguously} epistemic. To give an example, suppose we are certain that an experimenter has implemented one of two preparation procedures which are operationally distinct, hence represented by different states $S_1$ and $S_2$; but suppose that we simply \tit{do not know} which one was implemented (due perhaps to lack of communication from them, loss of records, amnesia, or anything else of that kind). Such situations inevitably arise in practice, and it can be useful to deal with them by including this additional strictly epistemic uncertainty in our formal reasoning. 

The method usually adopted in the literature for incorporating epistemic uncertainty is simply to `absorb' it into the definition of preparations, by treating a preparation procedure about which we have epistemic doubt as being itself a new preparation procedure. For instance, if I assign epistemic probability $\lambda$ to the proposition that state $S_1$ was prepared, and $(1-\lambda)$ that $S_2$ was prepared, then I could represent this formally by defining a new preparation procedure associated to a state $S_3$, such that
\eqn{
\trm{Pr}(k|S_3,T,M) &:=& \lambda \, \trm{Pr}(k|S_1,T,M) \nonumber \\
&+&  (1-\lambda) \, \trm{Pr}(k|S_2,T,M) \qquad \forall T,M \, . \nonumber \\
&&
}
Many authors find this technique convenient because it forces the state space to be convex, which can be very useful for proving theorems. However, this strategy can be misleading if one does not use it carefully\footnote{A recent exchange in the literature provides an excellent case in point~\cite{Masanes2019,STACEY_critique,MGM_reply}. Responding to a criticism, the authors of Ref.\ \cite{Masanes2019} clarify that the states $S$ in the operational theory they consider are `agnostic' about whether the associated preparation procedure includes uncertainty of a purely epistemic character. While there may be valid reasons for doing this depending on one's aims, we agree with Ref.\ \cite{STACEY_critique} that this is a non-trivial restriction on the operational theories, and should be stated explicitly as an assumption.}, because it formally conflates uncertainty that is obviously epistemic with uncertainty that might have a non-epistemic physical origin. This can be a problem if one's goal is to contemplate possible non-linear dynamics as is the case here (recall the discussion in Sec~\ref{sec:warmup}); it also makes it difficult to compare different interpretations of the state, as we aim to do in Sec~\ref{sec:epistemic}. 

To see why, note that since epistemic uncertainty concerns only the knowledge or beliefs of the experimenters, it is always presumed that we can eliminate it (at least in principle) by adopting suitably rigorous experimental protocols, such as keeping meticulous records, establishing trust and communication among the scientists, \tit{etc}. Therefore if we wish to investigate whether some uncertainty implied by $S$ might have an ontic rather than epistemic origin, we should define the states $S$ only with respect to those preparation procedures in which every effort has been made to eliminate epistemic uncertainty. \tit{If} some residual uncertainty remains after this process, then -- and only then -- can we reasonably entertain the hypothesis that it \tit{might} have an ontic explanation. Note that if we adopt this rule, then the mathematical space of states $S$ can indeed be non-convex. 

\subsubsection{Ensemble formalism}
In order to maintain a strict separation between uncertainty that is clearly epistemic (\tit{ie}.\ which we know how to eliminate) and that which is not, we cannot simply absorb the former kind of uncertainty into the preparation procedures. Instead, we shall formally represent such uncertainty using an \tit{ensemble}, which is a set $E := \{ \lambda_j : S_j \}_{j}$ where the index $j$ runs over a finite set of integers, $j \in \{ 1,2,\dots \}$. An ensemble refers to a situation of strictly epistemic uncertainty, in which we are sure that the correct state assignment is in the set $\{ S_j \}_j$ but we do not know which, and the $\{ \lambda_j \}_j$ represents our probabilities for each $S_j$ to be correct. We use the convention that a trivial ensemble consisting of only one possible state is simply referred to as a `state', and the term `ensemble' is reserved exclusively for cases where there is more than one possible state assignment. 

An ensemble of ensembles is again an ensemble. Formally, given ensembles $E_1 = \{ \lambda^{(1)}_k : S_k \}_k$ and $E_1 = \{ \lambda^{(2)}_j : S_j \}_j$, where the domains of indices $j$ and $k$ can in general overlap (ie.\ a state $S_k$ is common to both ensembles iff $k=j$), we might be `uncertain about our uncertainty' and assign probability $p$ that the ensemble is $E_1$ and $(1-p)$ that the ensemble is $E_2$. Intuitively, this should be equivalent to an ensemble that merely represents uncertainty over the union of the states in the two sub-ensembles, $S_i \in \{S_k \}_k \cup \{ S_j \}_j$, represented as $E_{1\cup2} := \{\lambda^{(3)}_i : S_i \}$, where the probability of $S_i$ is the sum of the probabilities for $S_i$ in each ensemble, weighted by the probabilities of the relevant ensemble, that is,
\eqn{
\lambda^{(3)}_i := p \lambda^{(1)}_i \,  \delta_{i,k} + (1-p) \lambda^{(2)}_i \, \delta_{i,j} \, .
}
Note that if we treat a state $S$ as equivalent to an ensemble of one element, $S \equiv \{1 : S \}$, this means that repeated elements in an ensemble can be formally combined into a single element, and in particular, $\{\lambda_k : S \}_k = \{1 : S \}$, which agrees with intuition.

Finally, to compute the probabilities of outcomes $k$ for some measurement $M$ performed on an ensemble $E := \{ \lambda_j : S_j \}_j$, we must compute the probabilities for the individual cases $S_j$ and then take the sum of these weighted by their respective epistemic likelihoods:
\eqn{
\trm{Pr}(k|E,M) := \zum{j}{} \, \lambda_j \trm{Pr}(k|S_j,M)  \, .
}

\subsubsection{Kinematical equivalence classes}

Consider two states $S_1$ and $S_2$ that yield the same probabilities for all measurements:
\eqn{
\trm{Pr}(k|S_1,M) = \trm{Pr}(k|S_2,M) \, \quad \forall M  \, .
}
In this case, we shall say that $S_1$ and $S_2$ belong to the same statistical \tit{equivalence class}, denoted $\mathcal{E}$. We emphasize that this equivalence class is defined only \tit{kinematically}, for as we will see later, some theories might permit transformations that map $S_1$ and $S_2$ to statistically non-equivalent final states.

More generally, two \tit{ensembles} that produce the same statistics for all measurements are also said to belong to the same equivalence class $\mathcal{E}$. That is, for any pair of ensembles in the same equivalence class $E_1,E_2 \in \mathcal{E}$ we have: 
\eqn{
\trm{Pr}(k|E_1,M) = \trm{Pr}(k|E_2,M) \quad \forall M  \, .
}
Since the probabilities $\trm{Pr}(k|E,M)$ are the same for any $E \in \mathcal{E}$, we shall sometimes refer to them using the notation $\trm{Pr}(k|\mathcal{E},M)$ to emphasize that the probabilities only depend on the equivalence class and not on the particular ensemble that generates them. 

It is important to note that individual states (i.e.\ which are not ensembles) can in general be statistically equivalent to ensembles. For instance, we can \tit{define} a state $S_{\mathcal{E}}$ such that it satisfies
\eqn{
\trm{Pr}(k|S_{\mathcal{E}},M) := \trm{Pr}(k|\mathcal{E},M)  \quad \forall M  \, 
}
for some equivalence class $\mathcal{E}$. Note that although the state $S_{\mathcal{E}}$ produces the same statistics as any ensemble $E \in \mathcal{E}$, they are conceptually distinct in two essential aspects, as we now explain.

Firstly, the state $S_{\mathcal{E}}$ -- like any state -- has a flexible interpretation: it may be taken as an \tit{ontic} state (describing a real state of affairs) or as an epistemic state (representing one's knowledge or beliefs). By contrast, as we explained earlier, the probabilities in an ensemble $E$ are \tit{always} epistemic by definition: they represent our uncertainty about which of the $\{S_j\}$ should be applied to the given situation.

Secondly, we are assuming that the theory imposes no restrictions on the possible ensembles, thus, any logically possible ensemble constructed by taking an arbitrary probabilistic mixture of states is assumed to be a valid ensemble in the theory. By contrast, not every logically possible state is considered to be allowed by the theory. In particular, there might exist ensembles $E \in \mathcal{E}$ whose statistics cannot be reproduced by any single state, i.e.\ while it is always logically possible to define the state $S_{\mathcal{E}}$ in the equivalence class $\mathcal{E}$, a given theory may exclude it from the set of valid states.

Finally, it will be useful to define an \tit{informationally-complete} measurement, denoted by the special symbol $\Pi$ with associated outcomes $i \in \{1,2,\dots \}$, as a measurement whose probabilities $p_S := \{ \trm{Pr}(i|S,\Pi) \}_i$ when measured on any state $S$ are sufficient to determine the probabilities that one would obtain by performing \tit{any other measurement} on $S$ instead. Formally, two states $S_1$ and $S_2$ yield the same probabilities for the measurement $\Pi$ (that is, $p_{S_1} = p_{S_2}$) if and only if they belong to the same equivalence class, that is:
\eqn{ \label{eqn:info-complete}
&& p_{S_1} = p_{S_2} \nonumber \\
&\Longleftrightarrow& \trm{Pr}(k|S_1,M) = \trm{Pr}(k|S_2,M) \, \quad \forall M  \, .
}
Such measurements therefore provide a useful way of indicating that two states belong to the same equivalence class, without needing to explicitly contemplate all possible measurements. In this work we restrict attention to operational theories which have at least one informationally complete measurement that can be performed on every system within the theory.

\subsubsection{Mixed states and superpositions}
Given the aforementioned differences between ensembles $E \in \mathcal{E}$ and states $S_{\mathcal{E}}$ that produce equivalent statistics, we shall adapt some familiar terminology to distinguish them: the former will be called \tit{proper mixtures} and the latter will be called \tit{improper mixed states}. Conversely, any state which does \tit{not} produce statistics equivalent to some ensemble will be called a \tit{pure state}\footnote{Equivalently, a state is called pure iff there is some measurement that `tests' for it, that is, a measurement that predicts a certain outcome with certainty only if the system is assigned that particular state.}. Our usage of these terms follows closely their conventional usage in the special case of quantum theory\footnote{There is a subtle difference: in some presentations one assumes that isolated systems cannot evolve dynamically from pure states into improper mixed states, in which case an `improper mixed state' always refers to a `reduced' state of a system that is correlated with some larger environment. Our more general framework does not impose this restriction; it accommodates improper mixed states not correlated to any environment.}.

Besides mixed states, we would also like to define a `superposition state' in our general operational framework. Note that the state of `being in superposition' is typically understood in contradistinction to some concept of `classical' states, these representing a special subset of states which are by definition \tit{not} superpositions. It is characteristic of such states that they represent situations in which a special subset of observable physical quantities can be assigned definite values -- these quantities being the ones that describe the `classical limit' of the theory. We shall presume that the same holds true in any of the generalized theories that we consider: given any system, we assume there is a special set of states $\{ S^{\trm{(cl)}}_k \}_k$ that play the role of the `classical states' of the theory. Since these are supposed to represent states in which a certain set of quantities (the classical observables) take definite mutually distinct values $k$, we shall assume there exists some measurement $M^{\trm{(cl)}}$ that is capable of mutually perfectly distinguishing among any subset from this set of states. This inspires the following definition:

\defn{\label{def:superposition}Let $S$ be a state of a system -- not an ensemble -- which has the following property: if subjected to the measurement $M^{\trm{(cl)}}$ (i.e.\ the measurement that distinguishes among the classical states of the system) the probabilities to obtain outcomes $k$ are non-trivial, that is, two or more values of $k$ are assigned non-negligible probabilities. Then we call such a state a \tit{quasi-superposition}.
}

The intuition behind this definition is the idea that the essence of a `superposition' lies in the combination of two factors (i) it does not yield definite values for `classical' variables and (ii) it is not an ensemble. The latter point means that, \tit{if} we adopt an ontic interpretation of states, then the indefiniteness of classical values represented by a quasi-superposition state must be due to a metaphysical indeterminacy at the ontological level: it cannot be explained away as representing merely our lack of knowledge of the true determinate values. 

Note that by this definition, a quasi-superposition state need not be a pure state -- hence the prefix `quasi'. Thus, even a theory that is operationally equivalent to classical theory can admit a notion of `quasi-superposition', in the form of some notion of `irreducible randomness', that is, in which the \tit{ontological} states of the theory include ones that have indeterminate classical values. 

\subsection{Main argument}

We contemplate the following scenario: a system labelled $\mathcal{S}_1$ is sealed inside a laboratory with a scientist, whom we model as another system $\mathcal{S}_2$. The whole laboratory and its contents (including the scientist), hereafter called the \tit{total system}, will be labelled $\mathcal{S}_{1,2}$. The task of the scientist $\mathcal{S}_2$ will be to determine -- purely by operations performed inside the lab -- whether or not the state of the total system $\mathcal{S}_{1,2}$ is in a `superposition' state or not.

We begin by noting that if $\mathcal{S}_{2}$ were able to succeed in their task by performing some `mystery operations' inside the lab, then it ought to be possible for a scientist situated outside the lab to model the whole process in terms of states, transformations and measurements performed on the total system $\mathcal{S}_{1,2}$; for if that were not possible, we could not say that the operations carried out by $\mathcal{S}_{2}$ were compatible with our physical theory. We shall therefore adopt the perspective of a scientist outside the lab for the duration of the analysis. 

From this standpoint, if we judge that a measurement $M_1$ has been done on system $\mathcal{S}_{1}$ and some outcome $k$ obtained, then we assign this system a state $S^{(k)}_1$. Similarly, there is a special measurement $M_2$ we can do upon the scientist $\mathcal{S}_{2}$, resulting in some outcome $k'$, which we interpret as `interrogating $\mathcal{S}_{2}$ about what they observed'; thus $k'$ has the same range as $k$. If we perform $M_2$ and obtain $k'$, we shall accordingly assign the scientist the state $S^{(k')}_2$, which we interpret as implying that they would report having observed outcome $k'$ if interrogated again. Thus if the state of $\mathcal{S}_{2}$ is $S^{(k)}_2$ and we measure $M_2$, we must be certain to obtain $k'=k$:
\eqn{ \label{eqn:M2-repeatability}
\trm{Pr}(k'|S^{(k)}_{2},M_2) = \delta_{k,k'}
}
Furthermore, this interpretation suggests that if we were to measure both $M_1$ on $\mathcal{S}_{1}$ and $M_2$ on $\mathcal{S}_{2}$ then we should expect to obtain the same outcomes for both (i.e.\ $k'=k$). We shall therefore assume that this is so.  

We can now ask what state we should assign to the total system, consistent with the preceding conditions. The answer depends on whether our uncertainty about $k$ prior to measuring $M_1 M_2$ should be modeled as epistemic ignorance, by assigning $\mathcal{S}_{1,2}$ an ensemble, or whether it should be modeled as (possibly) having an ontological origin, as represented by a quasi-superposition. We consider each option in turn.

According to option `Ensemble', operations which constitute a measurement for $\mathcal{S}_2$ also constitute a measurement for everyone outside the lab; therefore the fact that the lab is sealed and isolated is irrelevant except insofar as it prevents an outsider from knowing which value of $k$ was actually obtained. Consequently in this case we should model the total system as being in an ensemble:
\eqn{ \label{eqn:total_ensemble}
E_{1,2} :=  \{ \lambda_k : S^{(k)}_{1}S^{(k)}_{2} \}_k \, .
}

Alternatively, according to option `Superposition', the operations performed by $\mathcal{S}_2$ do not constitute a measurement for us on the outside, but rather a transformation of the total system which leads us to assign it some final joint state $S^{(+)}_{1,2}$. 

Note that $M_2$ perfectly discriminates between a set of states $\{ S^{(k)}_{2} \}_k$, which we interpret as the scientist $\mathcal{S}_2$ having observed different values of the variable $k$ -- it is therefore natural to consider these states a subset of the `classical' states of our theory. Moreover, applying $M_2$ to $S^{(+)}_{1,2}$ does not yield a definite value of $k$ -- hence this state is a quasi-superposition by Definition~\ref{def:superposition}.

We can now pose the key question: is there any experiment that $\mathcal{S}_{2}$ could do on $\mathcal{S}_{1}$ which would allow them to determine whether the total system is assigned the quasi-superposition state $S^{(+)}_{1,2}$ or the ensemble $E_{1,2}$? 

If $\mathcal{S}_{2}$ indeed has the power to perform such an experiment on $\mathcal{S}_{1}$, we outside the lab should be able to model this process explicitly as a transformation $\tilde{T}_1$ applied to system $\mathcal{S}_{1}$, followed by a measurement $\tilde{M}_1$, whose outcome statistics are different depending on whether the system is in a superposition or not. Conversely, if no such $\tilde{T}_1$, $\tilde{M}_1$ exist, then $\mathcal{S}_{2}$ cannot succeed at their task.

We now prove that if $\tilde{T}_1$ is trivial, no $\tilde{M}_1$ exists that would enable $\mathcal{S}_{2}$ to succeed at their task.

\thrm{\label{thrm:kinematical-linear}In the scenario described, there is no measurement on $\mathcal{S}_{1}$ whose statistics would allow us to differentiate between $S^{(+)}_{1,2}$ and $E_{1,2}$.}

\tit{Proof:} 

First note that if the total system is assigned the ensemble $E_{1,2}$ as defined in Eq.\ \eqref{eqn:total_ensemble}, then $\mathcal{S}_{1}$ taken alone is assigned the state $S^{(k)}_{1}$ with probability $\lambda_k$, that is, we assign it the `reduced ensemble' $\tilde{E}_1$ defined as:
\eqn{ \label{eqn:reduced_ensemble}
\tilde{E}_{1} :=  \{ \lambda_k : S^{(k)}_{1} \}_k \, .
}
Alternatively, if the total system is assigned the state $S^{(+)}_{1,2}$ then in order to know what to assign to $\mathcal{S}_{1}$ alone, we require some rule that tells us how to obtain its state from the total state $S^{(+)}_{1,2}$. In quantum theory this would be the partial trace, but in general it could be something else. Fortunately, the details of this rule turn out not to matter: it is sufficient for our purposes to note that since the total state $S^{(+)}_{1,2}$ is not an ensemble, it fully determines the reduced state of system $\mathcal{S}_{1}$, which therefore also is not an ensemble; hence we can denote it $\tilde{S}_1$.

Our result can now be proven if we can show that no measurement on $\mathcal{S}_{1}$ can distinguish the reduced state $\tilde{S}_1$ from the ensemble $\tilde{E}_1$. In order to prove this, we contemplate doing a pair of measurements $\Pi_1$ and $M_2$ on systems $\mathcal{S}_{1}$ and $\mathcal{S}_{2}$ respectively, where $\Pi_1$ is the \tit{informationally-complete} measurement (as defined in Sec~\ref{sec:framework}) with outcomes $i \in \{ 1,2,\dots \}$, and $M_2$ corresponds to ``checking which outcome $k$ was observed by $\mathcal{S}_{2}$" as defined earlier, with outcomes $k'$. We now compute the marginal probabilities to obtain $i$ for each of the two cases Ensemble and Superposition and prove that they are the same.

For Ensemble, we can obtain the probabilities of $i$ immediately, since:
\eqn{  \label{eqn:E-marginal}
\trm{Pr}(i|\tilde{E}_{1},\Pi_1) &=& \zum{k}{} \, \sigma_{k} \, \trm{Pr}(i|S^{(k)}_{1},\Pi_1) \nonumber \\
 &:=& \zum{k}{} \, \sigma_{k} \, p_k \, .
}
For Superposition, the systems $\mathcal{S}_{1,2}$ are assigned the joint state $S^{(+)}_{1,2}$. We first compute the joint probabilities $\trm{Pr}(i,k')$ for both measurements $\Pi_1$, $M_2$, and subsequently sum over the $k'$ to find the marginal probabilities for $i$. The joint probabilities are:
\eqn{
&& \trm{Pr}(i,k'|S^{(+)}_{1,2},\Pi_1,M_2) \nonumber \\
&& =  
\trm{Pr}(k'|S^{(+)}_{1,2},\Pi_1,M_2) \, \trm{Pr}(i|k',S^{(+)}_{1,2},\Pi_1,M_2) 
\nonumber \\
&& = 
\lambda_{k'} \, \trm{Pr}(i|S^{(k')}_{1}S^{(k')}_{2},\Pi_1) 
\nonumber \\
&& = 
\lambda_{k'} \, \trm{Pr}(i|S^{(k')}_{1},\Pi_1) 
\nonumber \\
&& = 
\lambda_k' \,  p_k' \, ,
}
where the first line follows from the identity $\trm{Pr}(A,B|C)=\trm{Pr}(B|C)\trm{Pr}(A|B,C)$, the second line follows from the definition of $S^{(+)}_{1,2}$, and we used \eqref{eqn:factorize-P} to get the third line. Using this result, we sum over the $k'$ to obtain the marginal probabilities for $i$:
\eqn{ \label{eqn:S-marginal}
\trm{Pr}(i|\tilde{S}_1,\Pi_1) &=& \zum{k'}{} \, \trm{Pr}(i,k'|S^{(+)}_{1,2},\Pi_1,M_2) \nonumber \\
&=&  \zum{k'}{} \, \lambda_k' \,  p_k' \, .
}
Comparing the results \eqref{eqn:E-marginal} and \eqref{eqn:S-marginal} and noting that $k'$ by definition has the same domain as $k$, we see that the reduced state $\tilde{S}_1$ produces the same probabilities for the measurement $\Pi_1$ as does the reduced ensemble $\tilde{E}_1$; but since $\Pi_1$ is informationally-complete, Eq.\ \eqref{eqn:info-complete} implies that:
\eqn{
\trm{Pr}(k|\tilde{S}_1,M) = \trm{Pr}(k|\tilde{E}_1,M) \, \quad \forall M \, ,
}
that is, they belong to the same equivalence class and \tit{no} measurement can distinguish them. $\Box$ \\

The preceding Theorem has the following implication: from a purely \tit{kinematical} point of view, if $\mathcal{S}_2$ is restricted to just performing operations that we model as measurements on $\mathcal{S}_1$, they cannot tell whether the total system is in the state $S^{(+)}_{1,2}$ or the ensemble $E_{1,2}$. As we saw, this stems from the fact that the improper mixture $\tilde{S}_1$ belongs to the same equivalence class as the proper mixture $\tilde{E}_1$.

We next consider the key question: is there some `mystery transformation' $T_1$ that would enable us to statistically distinguish between $\tilde{S}_1$ and $\tilde{E}_1$, thereby allowing $\mathcal{S}_2$ to succeed at their task? We will prove that the answer is `no' iff $T_1$ acts linearly on convex mixtures, as formalized by the following definition:

\defn{\label{def:convex-linear} Consider a transformation $T$ that maps the set of physical states $\mathcal{S}$ to itself\footnote{One might ask why we don't consider more general transformations that map physical states to \tit{ensembles} of physical states; we address this in Appendix~\ref{app:stochastic}.}. Given any state $S^{(k)}$, let $p_k$ be the vector of probabilities $\{ \trm{Pr}(i|S^{(k)},\Pi) \}_i$. Then the map $S^{(k)} \mapsto T(S^{(k)})$ induces a corresponding map $p_k \mapsto T(p_k)$, where 
\eqn{ 
T(p_k) := \{ \trm{Pr}(i|T(S^{(k)}),\Pi) \}_i \, .
}
The map $T$ is called \tit{convex-linear} if it acts linearly on arbitrary convex mixtures of $p_k$, that is if
\eqn{ \label{eqn:T-convex-linear} 
T\left( \zum{k}{} \, \lambda_k \, p_{k} \right) = \zum{k}{} \, \lambda_k \, T(p_{k}) \, ,
}
with $0 \leq \lambda_k \leq 1$ and $\sum_k \, \lambda_k = 1$.}

Armed with this definition, we now present our main theorem:

\thrm{\label{thrm:T-convex-linear} A transformation $T$ preserves the statistical equivalence of any state $S$ and ensemble $E$ that belong to the same equivalence class $\mathcal{E}$ iff $T$ is convex-linear as defined above.
}

\tit{Proof:} As usual, since an ensemble $E = \{ \lambda_k : S^{(k)} \}_k$ represents epistemic ignorance, we assume $T$ acts on it by acting individually on each of its components:
\eqn{ \label{eqn:T-ensemble}
T(E) :=  \{ \lambda_k : T(S^{(k)}) \}_k \, .
}
Let $S$ be a state that belongs to the same equivalence class $\mathcal{E}$ of $E$, and let $p_S$ (resp.\ $p_E$) be the vectors of probabilities generated by $S$ (resp.\ $E$) for the outcomes $i$ of the informationally-complete measurement $\Pi$. Since they belong to the same equivalence class, we must have $p_S = p_E$. This implies that $p_S$ can (like $p_E$) be decomposed as:
\eqn{ \label{eqn:T-improper}
p_S = \zum{k}{} \, \lambda_k \, p_{k} \, .
}
Using \eqref{eqn:T-ensemble}, if one performs $\Pi$ on the final ensemble $T(E)$, one obtains probabilities
\eqn{ \label{eqn:T-infocomplete}
T(p_E) &:=& \trm{Pr}(i|T(E),\Pi) \nonumber \\
 &=& \zum{k}{} \, \lambda_k \, \trm{Pr}(i|T(S^{(k)}),\Pi) \nonumber \\
&:=& \zum{k}{} \, \lambda_k \, T(p_{k})  \, .
}
Note that $T$ maps $S$ and $E$ to the same equivalence class iff $T(p_S) = T(p_E)$. Comparing \eqref{eqn:T-infocomplete} to \eqref{eqn:T-improper} we see that this occurs iff
\eqn{ 
T\left( \zum{k}{} \, \lambda_k \, p_{k} \right) = \zum{k}{} \, \lambda_k \, T(p_{k}) \, ,
}
or in other words, $T$ preserves the equivalence class structure (hence the statistical equivalence of $S$ and $E$) iff $T$ is convex-linear\footnote{At face value, this theorem might seem to contradict recent work on `quasi-linear dynamics'~\cite{Rembielinski_Caban_2020,Rembielinski_Caban_2021}, which describes a class of non-linear dynamical maps that are insensitive to different decompositions of the density matrix. In Appendix~\ref{app:quasilinear} we explain how quasi-linear dynamics are consistent with our result.}. $\Box$ \\

Putting together Theorems~\ref{thrm:kinematical-linear}~\&~\ref{thrm:T-convex-linear} we conclude that, in the general scenario described, it is impossible for the scientist in the closed laboratory to determine whether they are inside a superposition (that is, Local Definiteness holds) iff the transformations of the theory are convex-linear in the sense of Definition~\ref{def:convex-linear}.

There is one minor step needed to connect this general result to the quantum superposition principle, namely, we need to prove that when we restrict the kinematics of the theory to be that of quantum theory, then convex-linearity as defined in Definition~\ref{def:convex-linear} reduces to convex-linearity on density matrices as per Eq.\ \eqref{eqn:convex-linear-density}.\\

\prop{If we take the state space of our theory to be quantum, \tit{ie}.\ we identify the valid states with the set of density operators acting on a Hilbert space, and we identify valid measurements with POVMs defined on this space, then the transformations $T$ are convex-linear \tit{only if} they act linearly on density matrices.}

\tit{Proof:} Let the informationally-complete measurement $\Pi$ be associated to a POVM $\{ \Pi_i \}_i$; then every density operator $\rho$ is mapped to a unique set of probabilities via:
\eqn{ \label{eqn:Born}
\trm{Pr}(i|S,\Pi) &=& \tr{\rho_S \Pi_i} \, \nonumber \\
&:=& p_S \, .
}
Given a transformation $T$ defined on the density operators, the correspondence \eqref{eqn:Born} induces a corresponding map on their probability representations via
\eqn{
T(p_S) &:=& \{ \trm{Pr}(i|T(S),\Pi) \}_i   \nonumber \\
&=& \{ \, \tr{ T(\rho_S) \Pi_i}  \, \}_i \, .
}
From this definition it follows that, due to the linearity of the trace, if $T$ does \tit{not} act convex-linearly on the density operators it also cannot act convex-linearly on their probability representations. Equivalently, if $T$ is convex-linear on the probabilities as per Definition~\ref{def:convex-linear}, then it must be convex-linear on the density operators as per Definition~\ref{defn:convex-linear-density}. $\Box$

\section{Discussion: why epistemic? \label{sec:epistemic}}

We began this investigation by asking why no experiments to date have succeeded in detecting any deviation from the linear equations of quantum dynamics. Our foregoing results have proven that the question can be re-framed as follows: why do the laws of physics conform to the principle of Local Definiteness? If we can find a satisfying theoretical reason for why this principle is upheld by our most accurate experiments to date, then this would also explain why we have not observed any evidence of non-linear quantum dynamical effects.

Remarkably, if one adopts an epistemic interpretation of quantum states, then one is automatically committed to the principle of Local Definiteness. This was pointed out by Chris Fuchs during discussions at a recent workshop\footnote{The \tit{Wigner's Friends Theory Workshop}, San Francisco, December 2022; supported by Unitary Fund, FQXi, and Convergent Research.} devoted to Eugene Wigner's eponymous thought experiment~\cite{Wigner1995}. Fuchs emphasized that if the quantum state merely represents the knowledge or beliefs of someone outside the laboratory (that is, if the quantum state is interpreted as \tit{epistemic}) then it is \tit{obvious} that the friend inside the laboratory should be unable to tell the difference between being `in an entangled superposition' or not. Indeed, experiments within the lab should be powerless to reveal that the lab is in a superposition, just as experiments on a spinning roulette wheel are powerless to reveal whether someone has placed a bet on this particular spin of the wheel\footnote{Actually, the same argument leads to something much stronger than LD: local measurements on a system $S$ are unable to distinguish between \tit{any} pair of alternative states assigned to $S$ by another scientist, precisely because these `states' are not properties of the system but rather propositions about that other scientist's knowledge or beliefs. Local Definiteness is just a special case of this more general feature of epistemic interpretations.}.

This fact takes on a special significance in light of the foregoing results: for since LD implies the linearity of quantum dynamics, it follows that an epistemic interpretation of quantum states only makes sense if the dynamics is linear. We now argue that this explanation for why quantum dynamics is linear -- \tit{viz} that quantum states are epistemic -- is superior to alternative explanations that must be used if quantum states are regarded as ontic. To make our case, we identify two varieties of ontic interpretation, for which we shall make separate arguments in subsections~\ref{sec:abs_ontic} and \ref{sec:rel_ontic}. In Sec~\ref{sec:qbism} we discuss what our results imply for epistemic interpretations.

\subsection{Absolute ontic models \label{sec:abs_ontic}}

\defn{An interpretation is called \tit{absolute ontic} if every pair of distinct states $S$ and $S'$ of a given system imply two distinct and mutually exclusive ontological states of that system.}

This definition encompasses what Harrigan \& Spekkens call $\psi$-\tit{complete} ontological models, in which there is a one-to-one correspondence between quantum states and ontological states, as well as what they call $\psi$-\tit{supplemented} ontological models, where the quantum state is an incomplete specification of the ontological state and hence it is possible for distinct ontological states to be assigned the same quantum state~\cite{Harrigan_Spekkens}.

To see why epistemic interpretations are superior to absolute ontic interpretations, consider Leibniz' principle as recently elaborated by Spekkens~\cite{Spekkens_Leibniz}: 

\begin{quote}
    ``If an ontological theory implies the existence of two scenarios that are empirically indistinguishable in principle but ontologically distinct (where both the indistinguishability and distinctness are evaluated by the lights of the theory in question), then the ontological theory should be rejected and replaced with one relative to which the two scenarios are ontologically identical."
\end{quote}

Recall that the principle of Local Definiteness implies that it is empirically impossible to distinguish the reduced ensemble $\tilde{E}_1$ from the reduced state $\tilde{S}_1$ by any local operations. According to an absolute ontic interpretation, these situations represent two distinct ontological states of $\mathcal{S}_1$: in the former case the state is an ontological state corresponding to one of the $\{ S^{(k)}_1 \}_k$ (we just don't know which), whereas in the latter case it is in the ontological state associated to the improper mixed state $\tilde{S}_1$. 

By contrast, on an epistemic interpretation the states $\tilde{E}_1$ and $\tilde{S}_1$ are both only expressions of an external observer's knowledge or beliefs regarding the system, and so are compatible with the hypothesis that they refer to one and the same ontological state of affairs inside the laboratory.

This is a perfect use-case for Leibniz' principle: since $\tilde{E}_1$ and $\tilde{S}_1$ are \tit{empirically indistinguishable}, the principle says we ought to adopt an interpretation of quantum states according to which these states may refer to the \tit{same} local ontological state of $\mathcal{S}_1$. An epistemic interpretation permits this; an absolute ontic interpretation does not.

One might worry that we have only defeated a straw man, for there may be more sophisticated ways to interpret the quantum state as ontic. We next consider a prominent class of such interpretations.

\subsection{Relational ontic models \label{sec:rel_ontic} }

\defn{An interpretation is called \tit{relational ontic} if every pair of distinct states $S$ and $S'$ of a given system imply two distinct and mutually exclusive ontological \tit{relations} between that system and a designated external ``reference frame", denoted $F$. 
}

The motivating thought behind this class of ontic interpretations is that whenever we assign a quantum state to some system \tit{in practice}, we do so against the backdrop of the physical laboratory in which we happen to be situated, which determines (somehow) the relevant reference frame $F$. On this view, the state assigned to a system includes not only information about the ontic state of the system, but also information about its ontological relation to the implicit reference frame $F$. This means it is possible for a single ontological state of the system to be compatible with two distinct quantum states $S$ and $S'$, just in case this ontological state supports two distinct ontological \tit{relations} between itself and the frame $F$. 

As an analogy, suppose professor Bertlmann, who prefers to wear mis-matched socks, has some chance of wearing matching socks, say, when he is running late for a lecture and has dressed in a hurry. If on a given day his left sock is red, for example, then let us take the sock as being analogous to a physical system and `the sock is red' as analogous to an ontological fact about that system. This fact alone, however, does not fix the value of the ontological \tit{relation} between it and the other sock, which under the given conditions may be either `matched' or `mis-matched'. This illustrates how, in a relational ontic model, knowing the ontological state of a system does not fix its quantum state uniquely, unless we also specify the ontological \tit{relation} between the system and the frame $F$ in which it is being considered. 

It is noteworthy that relational ontic interpretations are not contemplated within the `ontological models' framework of Harrigan \& Spekkens. Some prominent interpretations do seem to fall within this category, including Healey's pragmatist interpretation\cite{Healey_2018,Healey_2021}, some readings of Rovelli's relational interpretation (RQM)\footnote{Rovelli interprets quantum states as equivalent to catalogs of probabilities, however since he remains agnostic about the interpretation of probability, it impossible to say whether he thereby endorses an ontic or epistemic reading of the quantum state~\cite{STACEY_RQM1}; indeed the literature contains examples of both~\cite{Dorato2020, RQM_Smerlak}. However, recent developments in RQM~\cite{DiBiagio_RQM,Adlam_RQM} have arguably forced it towards an ontic position~\cite{STACEY_RQM2,STACEY_RQM3}. Moreover, if we adopt Leifer's definition of ontic as ``$\dots$ something that objectively exists in the world [$\dots$] that would still exist if all intelligent beings were suddenly wiped out from the universe", then the quantum state in RQM certainly qualifies.}~\cite{ROVELLI_96,SEP_RQM,Dorato2020}, and perhaps the `Contexts, Systems and Modalities' (CSM) interpretation of Auff{\`e}ves and Grangier~\cite{Auffeves2020,Auffeves_2022}. 

Unlike absolute ontic interpretations, relational ontic interpretations are not susceptible to the argument from Leibniz' principle, because they allow two distinct quantum states to be associated with a single ontological state. Thus, the scientist's inability to distinguish $\tilde{E}_1$ from $\tilde{S}_1$ could possibly be accounted for by the fact that, so long as they are confined inside their laboratory, they do not have access to information about the extrinsic relations between their laboratory and the `outer laboratory' relative to which the state of the total system $\mathcal{S}_{1,2}$ is defined.

However, there is another principle for evaluating the explanatory power of hypotheses according to which the relational ontic explanation is inferior to the epistemic explanation. Given competing hypotheses $A$ and $B$ which both purport to explain some empirical phenomenon $X$, we shall judge an explanation superior if it is \tit{less able} to explain a \tit{violation} of $X$. The intuition here is that a good explanation of $X$ is one that would be falsified (or very nearly so) if it were to transpire that $X$ does not occur after all. In the present example, we should ask: if future experiments were to find evidence of non-linear quantum dynamical effects, how easily could this be accounted for by an epistemic versus a relational ontic interpretation? 

It is straightforward to see that the relational ontic interpretation would easily survive this hypothetical experimental development; for in any relational theory of mechanics, it is practically a rule that local phenomena of a system at one time depend on the extrinsic relations of that system to other systems at some earlier time. For example, consider the instantaneous relative distance (as defined in some conventionally chosen reference frame) between the Earth and the Sun. While it is true that I cannot determine this distance by any local experiment performed on Earth, I can determine what it was in the past, for instance by measuring the local temperature. 

Thus, if non-linear dynamics were someday discovered, allowing us to tell the difference between $\tilde{E}_1$ and $\tilde{S}_1$ by local experiments on $\mathcal{S}_1$ in our thought experiment, a relational-ontic interpretation could explain this without too much trouble by saying that the local experiment is merely revealing the difference between two ontic relations, say $R_E$ and $R_S$ respectively, defined between the local laboratory and some extrinsic frame $F$, as they stood at some point in the causal past of the experiment.

Now consider how an epistemic interpretation would fare in light of such a discovery. Since the difference between $\tilde{E}_1$ and $\tilde{S}_1$ is in this case not a \tit{physical} difference, but instead represents a difference in an external observer's knowledge or beliefs about the laboratory and its contents, then it would be deeply disturbing to find that this difference could be detected by experiments within the laboratory. It would imply that some hitherto unknown force has breached the laboratory walls\footnote{Admittedly, this alone would not be entirely unprecedented -- just think of the discovery of X-rays.}, but more alarmingly, that this force conveys very specific information about other people's mental states of knowledge or beliefs. To insist on an epistemic interpretation under these circumstances, in other words, would be tantamount to endorsing mental telepathy as the explanation. Even if one allows that there might be some phenomena for which telepathy is a plausible explanation, this is not one of them, because for this hypothetical case there is a perfectly sound physical explanation: the quantum state is ontic and its dynamics is non-linear. Putting it more bluntly, faced with compelling experimental evidence for non-linear dynamics, epistemic interpretations would be stone dead.

We conclude that while both epistemic and relational ontic interpretations can explain why the principle of Local Definiteness should hold (hence too the linearity of quantum dynamics), only the epistemic interpretation stands the risk of being falsified if experiments were to find evidence of non-linear effects; it is therefore the better explanation of the two.

\subsection{What ontology? \label{sec:qbism}}

As already noted, according to epistemic interpretations, nothing prevents us from asserting that the real situation inside the laboratory is the same, regardless of whether or not it an ensemble or a superposition from the outside. A skeptic might therefore challenge us to specify just what remains of the quantum formalism that could possibly constitute this ``real situation'', according to an epistemic account. As this question can only be answered in the context of a specific epistemic interpretation, a full answer lies beyond the scope of this article; we can however make some preliminary remarks. 

One option is to adopt a \tit{psi-epistemic} ontological model, according to which the state $S$ represents partial epistemic knowledge about some \tit{hidden variables} that describe the true state of reality inside the laboratory~\cite{Harrigan_Spekkens}. However, a number of no-go theorems have shown that psi-epistemic ontological models cannot reproduce operational quantum theory in general, and they have therefore fallen out of favour (see Ref.\ \cite{Leifer_review} for a review).

The alternative is to adopt an interpretation in which quantum states are epistemic, but where there are no hidden variables. There are very few interpretations with this feature; two notable examples are the information-theoretic interpretation of Bub and Pitowsky~\cite{BUB_PITOWSKI, BUB_19}, and the subjective Bayesian interpretation known as QBism (see \tit{eg}.\ Refs~\cite{Fuchs10a,QBism_FDR2014,FuchsStacey2018}). In both cases, a quantum state does not represent a state of affairs in reality (not even partially); instead the quantum state is regarded as equivalent to a catalog of probability assignments for the outcomes of possible measurements that could be performed on the system. The interpretations differ as to whether the probabilities are `objective' or `subjective', but they agree that probabilities (hence quantum states) are epistemic: they reflect knowledge or beliefs about anticipated measurement outcomes. It is then clear that the quantum state in such interpretations is \tit{fully} epistemic; by itself it says nothing about the real situation inside the lab. But if the quantum state has no direct hold upon reality, then what does?

The first thing to note is that, within the epistemic interpretations under discussion, so long as the lab remains `sealed', that is, so long as we do not `interrogate' the scientist $\mathcal{S}_{2}$ as to what value of $k$ they observed, we on the outside \tit{cannot make any claim} about the `real situation' inside the lab. Insofar as such a `real situation' can be said to exist at all, it can only exist for the scientist $\mathcal{S}_{2}$ encapsulated in the lab\footnote{This conclusion follows from arguments laid out in detail in Refs.\ \cite{BONG_2020,Cavalcanti_Bubble,Cavalcanti_Qcausality_2021}; it amounts to rejecting the assumption of `Absoluteness of Observed Events'.}.  

Consequently, to discuss the `real situation' inside the lab, we must break a rule so far respected within this article and speculate about the point of view of the encapsulated scientist $\mathcal{S}_{2}$. If we do so, we can gain a clue as to what the `real situation' might be for them, by recalling the definition of Local Definiteness: \tit{a scientist confined to an enclosed laboratory cannot determine by any local experiment whether the laboratory and all its contents are in a definite effectively classical state, or in a superposition of such states}. 

Within the epistemic interpretations we are considering, this implies that if $\mathcal{S}_2$ observes some outcome $k$ (corresponding to some effectively classical state of affairs inside the lab), they can safely assume that -- at minimum -- this outcome $k$ is part of the invariant `real situation' pertaining to the inside the lab. Somebody outside the lab could still assign it a superposition state, and they might have very good reasons for doing so -- but this information is irrelevant to the encapsulated scientist, so long as their concerns remain localized to the interior\footnote{One is reminded of the similar role played by the Principle of Relativity, which guarantees that a scientist inside an inertially moving laboratory can safely assume that all systems in the laboratory are stationary, so long as they restrict attention to experiments confined to the interior of the lab.}. 

Thus, if there is to be a non-trivial ontology implied by an epistemic interpretation, our results suggest that it has the following minimal features: (i) elements of reality are associated to measurement outcomes; (ii) the latter are only `locally real', in the sense that they are real relative to the observer or observers who measure them; (iii) due to Local Definiteness, the local reality of some outcome $k$ is secured for all operations that can be performed `locally', that is, within the encapsulated laboratory.   

\section{Conclusion}

We have proven that for a wide class of physical theories, a scientist in an enclosed laboratory cannot determine whether they are inside a superposition or not, iff the dynamics of the theory is convex-linear. Consequently, if we choose to adopt the principle of Local Definiteness, then linearity of the dynamics follows automatically. 

We have furthermore argued that if quantum states are ontic, then there is no strong reason to expect Local Definiteness to be true, except insofar as it is derivable from assumptions about the dynamics. By contrast, if quantum states are epistemic, then we can only reasonably consider theories that satisfy Local Definiteness. Hence epistemic interpretations can better explain why quantum dynamics has so far been observed to be linear.

We now mention some limitations of our argument and some questions that remain open. 

Firstly, in Sec~\ref{sec:qbism} we argued that the local real situation inside the lab can be identified with the measurement outcomes obtained by the scientist in the lab. However, this opens up many more questions: is this local situation in principle accessible to observers outside? If so, how and under what conditions? 

Secondly, our argument makes a formal distinction between kinematics and dynamics throughout, but this is somewhat ironic since Spekkens has elsewhere applied Leibniz' principle to argue that this distinction is unfounded~\cite{spekkens_paradigm_2015}. Thus, if our argument really \tit{requires} us to make a formal split between kinematics and dynamics, our appeal to Leibniz' principle might actually undermine our argument. It would therefore be desirable to prove that our results still hold in a framework that does not separate kinematics from dynamics.

Finally, we have here only focused on one aspect of quantum theory, namely the linearity of the dynamics. However, a full evaluation of different interpretations should consider their ability to explain a wider range of quantum phenomena. The present results provide one more reason to take epistemic interpretations seriously, but much work still remains to show that they are decisively better than ontic alternatives.

\acknowledgements

This publication was made possible through the support of Grant 62424 from the John Templeton Foundation. The opinions expressed in this publication are those of the author and do not necessarily reflect the views of the John Templeton Foundation. Thanks to Chris Fuchs, Blake Stacey, Matt Weiss, and Thomas Galley for feedback that led to improvements in the article.

\begin{appendix}

\section{What about stochastic non-linear dynamics? \label{app:stochastic}}
In our proof of Theorem~\ref{thrm:T-convex-linear} we considered transformations $T$ that map physical states to (unique) physical states. One might worry that such maps exclude so-called \tit{stochastic non-linear maps}, which include important classes of non-linear dynamical theories such as the Ghirardi–Rimini–Weber~\cite{GRW} and Continuous Spontaneous Localization~\cite{Pearle_CSL} models (for recent overviews see Refs~\cite{collapse_SEP,Bassi_review,Hejlesen_thesis}). There is no reason for concern: our framework is general enough to encompass stochastic non-linear dynamics, as we now explain.

A transformation $T$ will be called \tit{stochastic} in our framework if it transforms a pure state $S$ into a state which is not pure, that is, either into an improper mixed state (recall Sec~\ref{sec:framework}) or into an ensemble of states. The former case is explicitly included within the scope of transformations contemplated by Theorem~\ref{thrm:T-convex-linear}, so we need only concern ourselves with the latter case. Note that transformations of this kind act on a state $S$ by transforming it into an ensemble, which can be written as:
\eqn{ \label{eqn:ensemble-stochastic}
T(S) = \{\lambda_k : T_k(S) \}_k \, ,
}
where the component transformations $T_k$ may be assumed without loss of generality to be transformations in the ordinary sense, that is, they map physical states to unique physical states. 

By definition, the ensemble \eqref{eqn:ensemble-stochastic} represents a situation of epistemic ignorance: it implies one of the transformations $T_k$ has actually occurred, we just don't know which one. Note that this means each $T_k$ individually must be an allowed transformation of the theory, one that could in principle be implemented on its own. 

Now recall that Theorem~\ref{thrm:T-convex-linear} says that a transformation $T$ is convex-linear iff it does not permit us to distinguish any state $S$ and ensemble $E$ belonging to the same equivalence class $\mathcal{E}$. Note that if there exists a valid transformation $T_{k'}$ that allows us to distinguish $S$ from $E$, we can always contrive a valid stochastic transformation of the form \eqref{eqn:ensemble-stochastic} containing $T_{k'}$ that retains this power -- for instance, one in which the probability of implementing any other $T_k$ with $k \neq k'$ is arbitrarily small. Conversely, if each $T_k$ individually cannot enable us to distinguish $S$ from $E$, then applying a stochastic transformation that merely represents ignorance about which of the $T_k$ was implemented can hardly improve our chances. 

In conclusion, stochastic transformations of the form \eqref{eqn:ensemble-stochastic} can distinguish $S$ from $E$ iff there is at least one valid map $T_k$ that can do so. It then follows by Theorem~\ref{thrm:T-convex-linear} that such stochastic transformations are powerless to differentiate $S$ from $E$ iff all valid transformations are convex-linear.

\section{What about quasi-linear dynamics? \label{app:quasilinear}}
A special class of non-linear quantum dynamical transformations has recently been proposed, called \tit{quasi-linear} dynamics~\cite{Rembielinski_Caban_2020,Rembielinski_Caban_2021}. Formally, a transformation $T^{\trm{QL}}$ is called quasi-linear iff it satisfies the following two properties. First, it preserves the convex structure of density operators:
\eqn{ \label{eqn:quasilinear1}
&& T^{\trm{QL}}\left( \zum{k}{} \, \lambda_k \rho_k \right) \nonumber \\
&& = \zum{k}{} \, \tilde{\lambda}_k T^{\trm{QL}}(\rho_k)
}
where $\tilde{\lambda}_k >0$, $\lambda_k >0$, and $\sum_k \tilde{\lambda}_k = \sum_k \lambda_k = 1$. Secondly, the final state $T^{\trm{QL}}(\rho)$ does not depend on which decomposition of $\rho$ is used to compute it. That is, if we have two different convex decompositions of $\rho$,
\eqn{
\rho &=& \zum{k}{} \, \lambda^{(1)}_k \rho^{(1)}_k \label{eqn:decomp1} \\
&=& \zum{j}{} \, \lambda^{(2)}_j \rho^{(2)}_j \, \label{eqn:decomp2},
}
then we must have
\eqn{ \label{eqn:quasilinear2}
T^{\trm{QL}}(\rho) &=& \zum{k}{} \, \tilde{\lambda}^{(1)}_k T^{\trm{QL}}(\rho^{(1)}_k) \nonumber \\
&=& \zum{j}{} \, \tilde{\lambda}^{(2)}_j T^{\trm{QL}}(\rho^{(2)}_j) \, .
}
Note that $T^{\trm{QL}}$ reduces to a linear transformation of density operators iff $\tilde{\lambda}_k = \lambda_k$ in \eqref{eqn:quasilinear1}.

The significance of this for our present work is that it is tempting to interpret the two convex decompositions \eqref{eqn:decomp1} and \eqref{eqn:decomp2} as representing different epistemic ensembles 
$E_1 := \{ \lambda^{(1)}_k : S^{(1)}_k \}_k$ and $E_2 := \{ \lambda^{(2)}_j : S^{(2)}_j \}_j$ respectively, where the measurement statistics of the states $S^{(1)}_k$, $S^{(2)}_j$ are computed using the density matrices $\rho^{(1)}_k$, $\rho^{(2)}_j$ respectively. The problem is that if we interpret the convex decompositions in this way, then it is clear that (i) the epistemic ensembles belong to the same statistical equivalence class (since they have the same total density matrix) and (ii) they cannot be distinguished by any quasi-linear transformation. But this would contradict our Theorem~\ref{thrm:T-convex-linear}, since the latter implies that transformations which cannot distinguish ensembles in the same equivalence class must be convex-linear, whereas quasi-linear transformations are \tit{non}-convex-linear whenever $\tilde{\lambda}_k \neq \lambda_k$. Something is clearly amiss!

To explain why there is no contradiction, we must stress that the word `ensemble' is often abused in standard quantum theory, being used on one hand to refer to improper mixed states, and on the other hand to refer to `proper mixtures' -- that is, representations of \tit{epistemic uncertainty} about which state actually obtains, which in our framework are exclusively called `ensembles'.

The reason why conflating these two notions does not normally lead to trouble is that standard quantum dynamics is linear, which makes it operationally impossible to distinguish an improper mixed state from an epistemic ensemble of states whose statistics are given by the same density matrix. The ontological difference between the two concepts is therefore normally operationally inconsequential and can be ignored.

The situation is drastically different when one is contemplating the possibility of non-linear quantum dynamics, as we explained in Sec~\ref{sec:warmup}. In this case it is absolutely essential to make a formal distinction between proper and improper mixtures (see \tit{eg}.\ remarks in Refs~\cite{Hejlesen_thesis,Pienaar_thesis,JORDAN_1993,Haag1978} ). To avoid inconsistency, we must be explicit about how the mixed density matrices are to be interpreted in the defining equations \eqref{eqn:quasilinear1} \eqref{eqn:quasilinear2} of quasi-linear dynamics.

In Ref~\cite{Rembielinski_Caban_2020} the two decompositions \eqref{eqn:decomp1} and \eqref{eqn:decomp2} are referred to as different `ensembles', but it is not specified whether this means they represent distinct \tit{epistemic} mixtures of states, or whether each one simply refers to an improper mixed state at the ontological level. Fortunately we can fill in this gap ourselves, for a little thought shows that the probability weights $\tilde{\lambda}_k,\lambda_k$ in the definition \eqref{eqn:quasilinear1} \tit{cannot} after all be interpreted as representing epistemic ignorance. The reason is that this would imply that $T^{\trm{QL}}$ acts on an ensemble $E := \{ \lambda_k : S_k \}_k$ as follows:
\eqn{
T^{\trm{QL}}(E) := \{ \tilde{\lambda_k} : T^{\trm{QL}}(S_k) \}_k \, ,
}
where $\tilde{\lambda_k} \neq \lambda_k$, that is, where $T^{\trm{QL}}$ actively changes the probability weights. If the latter are epistemic, then unless $T^{\trm{QL}}$ somehow acts on our brains to change our beliefs in subtle ways, this should not happen. In different words, if $\lambda_k$ merely represents our \tit{epistemic} uncertainty as to whether the state really is $S_k$ prior to the transformation, then simply knowing that we have applied $T^{\trm{QL}}$ to the system should not make us more or less certain that the state we applied it to was in fact $S_k$; after the transformation the epistemic weight for the state to be $T^{\trm{QL}}(S_k)$ should still be just $\lambda_k$ (Indeed we explicitly assumed this in the first step \eqref{eqn:T-ensemble} of our proof of Theorem~\ref{thrm:T-convex-linear}). 

We therefore see that the premise leading to an apparent contradiction with our framework was false: we \tit{cannot} after all interpret the two convex decompositions \eqref{eqn:decomp1} and \eqref{eqn:decomp2} as representing different epistemic ensembles $E_1$ and $E_2$. Instead, within our framework, these two decompositions must represent two different improper mixed states $S_1$, $S_2$, which can each be thought of as being specified by the density matrix $\rho$ together with a choice of parameters (either $\lambda^{(1)}_k$ or $\lambda^{(2)}_j$) that specify the particular decomposition of $\rho$ being considered. 

A general non-linear transformation $T$ as defined in our framework can therefore in principle distinguish between $S_1$ and $S_2$ by being sensitive to these parameters -- for instance we could consider maps that obey \eqref{eqn:quasilinear1} relative to the decomposition associated to the state, but which violate \eqref{eqn:quasilinear2} by transforming states with initially the same $\rho$ into states associated with distinguishable density matrices.
Understood in this way, we see that the quasi-linear maps are the non-trivial subset of non-linear maps for which states having the same density matrix are mapped to states having the same density matrix, even though their other ontological parameters (specifying their decompositions) may change. 

Having thus brought quasi-linear transformations into concordance with our framework, we see that Theorem~\ref{thrm:T-convex-linear} must apply to them as well: a quasi-linear transformation $T^{\trm{QL}}$ cannot distinguish a state $S$ from an epistemic ensemble $E$ belonging to the same equivalence class iff $T^{\trm{QL}}$ is convex-linear, that is, $\tilde{\lambda}_k = \lambda_k$ in \eqref{eqn:quasilinear1}. 

\end{appendix}


\begin{thebibliography}{48}%
\makeatletter
\providecommand \@ifxundefined [1]{%
 \@ifx{#1\undefined}
}%
\providecommand \@ifnum [1]{%
 \ifnum #1\expandafter \@firstoftwo
 \else \expandafter \@secondoftwo
 \fi
}%
\providecommand \@ifx [1]{%
 \ifx #1\expandafter \@firstoftwo
 \else \expandafter \@secondoftwo
 \fi
}%
\providecommand \natexlab [1]{#1}%
\providecommand \enquote  [1]{``#1''}%
\providecommand \bibnamefont  [1]{#1}%
\providecommand \bibfnamefont [1]{#1}%
\providecommand \citenamefont [1]{#1}%
\providecommand \href@noop [0]{\@secondoftwo}%
\providecommand \href [0]{\begingroup \@sanitize@url \@href}%
\providecommand \@href[1]{\@@startlink{#1}\@@href}%
\providecommand \@@href[1]{\endgroup#1\@@endlink}%
\providecommand \@sanitize@url [0]{\catcode `\\12\catcode `\$12\catcode
  `\&12\catcode `\#12\catcode `\^12\catcode `\_12\catcode `\%12\relax}%
\providecommand \@@startlink[1]{}%
\providecommand \@@endlink[0]{}%
\providecommand \url  [0]{\begingroup\@sanitize@url \@url }%
\providecommand \@url [1]{\endgroup\@href {#1}{\urlprefix }}%
\providecommand \urlprefix  [0]{URL }%
\providecommand \Eprint [0]{\href }%
\providecommand \doibase [0]{https://doi.org/}%
\providecommand \selectlanguage [0]{\@gobble}%
\providecommand \bibinfo  [0]{\@secondoftwo}%
\providecommand \bibfield  [0]{\@secondoftwo}%
\providecommand \translation [1]{[#1]}%
\providecommand \BibitemOpen [0]{}%
\providecommand \bibitemStop [0]{}%
\providecommand \bibitemNoStop [0]{.\EOS\space}%
\providecommand \EOS [0]{\spacefactor3000\relax}%
\providecommand \BibitemShut  [1]{\csname bibitem#1\endcsname}%
\let\auto@bib@innerbib\@empty
\bibitem [{\citenamefont {Bassi}\ \emph {et~al.}(2013)\citenamefont {Bassi},
  \citenamefont {Lochan}, \citenamefont {Satin}, \citenamefont {Singh},\ and\
  \citenamefont {Ulbricht}}]{Bassi_review}%
  \BibitemOpen
  \bibfield  {author} {\bibinfo {author} {\bibfnamefont {A.}~\bibnamefont
  {Bassi}}, \bibinfo {author} {\bibfnamefont {K.}~\bibnamefont {Lochan}},
  \bibinfo {author} {\bibfnamefont {S.}~\bibnamefont {Satin}}, \bibinfo
  {author} {\bibfnamefont {T.~P.}\ \bibnamefont {Singh}},\ and\ \bibinfo
  {author} {\bibfnamefont {H.}~\bibnamefont {Ulbricht}},\ }\bibfield  {title}
  {\bibinfo {title} {Models of wave-function collapse, underlying theories, and
  experimental tests},\ }\href {https://doi.org/10.1103/RevModPhys.85.471}
  {\bibfield  {journal} {\bibinfo  {journal} {Rev. Mod. Phys.}\ }\textbf
  {\bibinfo {volume} {85}},\ \bibinfo {pages} {471} (\bibinfo {year}
  {2013})}\BibitemShut {NoStop}%
\bibitem [{\citenamefont {Ghirardi}\ and\ \citenamefont
  {Bassi}(2020)}]{collapse_SEP}%
  \BibitemOpen
  \bibfield  {author} {\bibinfo {author} {\bibfnamefont {G.}~\bibnamefont
  {Ghirardi}}\ and\ \bibinfo {author} {\bibfnamefont {A.}~\bibnamefont
  {Bassi}},\ }\bibfield  {title} {\bibinfo {title} {{Collapse Theories}},\ }in\
  \href@noop {} {\emph {\bibinfo {booktitle} {The {Stanford} Encyclopedia of
  Philosophy}}},\ \bibinfo {editor} {edited by\ \bibinfo {editor}
  {\bibfnamefont {E.~N.}\ \bibnamefont {Zalta}}}\ (\bibinfo  {publisher}
  {Metaphysics Research Lab, Stanford University},\ \bibinfo {year} {2020})\
  \bibinfo {edition} {summer 2020}\ ed.\BibitemShut {Stop}%
\bibitem [{\citenamefont {Hejlesen}(2019)}]{Hejlesen_thesis}%
  \BibitemOpen
  \bibfield  {author} {\bibinfo {author} {\bibfnamefont {Z.~L.}\ \bibnamefont
  {Hejlesen}},\ }\emph {\bibinfo {title} {{Nonlinear Quantum Mechanics}}},\
  \href
  {https://www.duo.uio.no/bitstream/handle/10852/70439/1/Masteroppgave_Laberg.pdf}
  {Master's thesis},\ \bibinfo  {school} {{University of Oslo}} (\bibinfo
  {year} {2019})\BibitemShut {NoStop}%
\bibitem [{\citenamefont {Pienaar}(2013)}]{Pienaar_thesis}%
  \BibitemOpen
  \bibfield  {author} {\bibinfo {author} {\bibfnamefont {J.}~\bibnamefont
  {Pienaar}},\ }\emph {\bibinfo {title} {{Causality Violation and Nonlinear
  Quantum Mechanics}}},\ \href {http://arxiv.org/abs/1401.0167} {Ph.D.
  thesis},\ \bibinfo  {school} {The University of Queensland} (\bibinfo {year}
  {2013})\BibitemShut {NoStop}%
\bibitem [{\citenamefont {Kent}(2021)}]{Kent_2021}%
  \BibitemOpen
  \bibfield  {author} {\bibinfo {author} {\bibfnamefont {A.}~\bibnamefont
  {Kent}},\ }\bibfield  {title} {\bibinfo {title} {Quantum state readout,
  collapses, probes, and signals},\ }\href
  {https://doi.org/10.1103/PhysRevD.103.064061} {\bibfield  {journal} {\bibinfo
   {journal} {Phys. Rev. D}\ }\textbf {\bibinfo {volume} {103}},\ \bibinfo
  {pages} {064061} (\bibinfo {year} {2021})}\BibitemShut {NoStop}%
\bibitem [{\citenamefont {Everett}(1957)}]{Everett57}%
  \BibitemOpen
  \bibfield  {author} {\bibinfo {author} {\bibfnamefont {H.}~\bibnamefont
  {Everett}},\ }\bibfield  {title} {\bibinfo {title} {{"Relative State"
  Formulation of Quantum Mechanics}},\ }\href
  {https://doi.org/10.1103/RevModPhys.29.454} {\bibfield  {journal} {\bibinfo
  {journal} {Rev. Mod. Phys.}\ }\textbf {\bibinfo {volume} {29}},\ \bibinfo
  {pages} {454} (\bibinfo {year} {1957})}\BibitemShut {NoStop}%
\bibitem [{\citenamefont {Barrett}(2018)}]{Barrett_SEP}%
  \BibitemOpen
  \bibfield  {author} {\bibinfo {author} {\bibfnamefont {J.}~\bibnamefont
  {Barrett}},\ }\bibfield  {title} {\bibinfo {title} {{Everett’s
  Relative-State Formulation of Quantum Mechanics}},\ }in\ \href@noop {} {\emph
  {\bibinfo {booktitle} {The {Stanford} Encyclopedia of Philosophy}}},\
  \bibinfo {editor} {edited by\ \bibinfo {editor} {\bibfnamefont {E.~N.}\
  \bibnamefont {Zalta}}}\ (\bibinfo  {publisher} {Metaphysics Research Lab,
  Stanford University},\ \bibinfo {year} {2018})\ \bibinfo {edition} {winter
  2018}\ ed.\BibitemShut {Stop}%
\bibitem [{\citenamefont {d'Espagnat}(1976)}]{dEspagnat}%
  \BibitemOpen
  \bibfield  {author} {\bibinfo {author} {\bibfnamefont {B.}~\bibnamefont
  {d'Espagnat}},\ }\href@noop {} {\emph {\bibinfo {title} {{Conceptual
  Foundations of Quantum Mechanics}}}}\ (\bibinfo  {publisher}
  {Addison-Wesley},\ \bibinfo {year} {1976})\BibitemShut {NoStop}%
\bibitem [{\citenamefont {Timpson}\ and\ \citenamefont
  {Brown}(2004)}]{Timpson_Brown_2004}%
  \BibitemOpen
  \bibfield  {author} {\bibinfo {author} {\bibfnamefont {C.~G.}\ \bibnamefont
  {Timpson}}\ and\ \bibinfo {author} {\bibfnamefont {H.}~\bibnamefont
  {Brown}},\ }\href {http://philsci-archive.pitt.edu/1616/} {\bibinfo {title}
  {Proper and improper separability}} (\bibinfo {year} {2004})\BibitemShut
  {NoStop}%
\bibitem [{\citenamefont {Baumann}\ and\ \citenamefont
  {Brukner}(2020)}]{Baumann2020}%
  \BibitemOpen
  \bibfield  {author} {\bibinfo {author} {\bibfnamefont {V.}~\bibnamefont
  {Baumann}}\ and\ \bibinfo {author} {\bibfnamefont {{\v{C}}.}~\bibnamefont
  {Brukner}},\ }\bibinfo {title} {Wigner's friend as a rational agent},\ in\
  \href {https://doi.org/10.1007/978-3-030-34316-3_4} {\emph {\bibinfo
  {booktitle} {Quantum, Probability, Logic: The Work and Influence of Itamar
  Pitowsky}}},\ \bibinfo {editor} {edited by\ \bibinfo {editor} {\bibfnamefont
  {M.}~\bibnamefont {Hemmo}}\ and\ \bibinfo {editor} {\bibfnamefont
  {O.}~\bibnamefont {Shenker}}}\ (\bibinfo  {publisher} {Springer International
  Publishing},\ \bibinfo {address} {Cham},\ \bibinfo {year} {2020})\ pp.\
  \bibinfo {pages} {91--99}\BibitemShut {NoStop}%
\bibitem [{\citenamefont {Rembieli{\'{n}}ski}\ and\ \citenamefont
  {Caban}(2020)}]{Rembielinski_Caban_2020}%
  \BibitemOpen
  \bibfield  {author} {\bibinfo {author} {\bibfnamefont {J.}~\bibnamefont
  {Rembieli{\'{n}}ski}}\ and\ \bibinfo {author} {\bibfnamefont
  {P.}~\bibnamefont {Caban}},\ }\bibfield  {title} {\bibinfo {title} {Nonlinear
  evolution and signaling},\ }\href
  {https://doi.org/10.1103/PhysRevResearch.2.012027} {\bibfield  {journal}
  {\bibinfo  {journal} {Phys. Rev. Res.}\ }\textbf {\bibinfo {volume} {2}},\
  \bibinfo {pages} {012027} (\bibinfo {year} {2020})}\BibitemShut {NoStop}%
\bibitem [{\citenamefont {Rembieli{\'{n}}ski}\ and\ \citenamefont
  {Caban}(2021)}]{Rembielinski_Caban_2021}%
  \BibitemOpen
  \bibfield  {author} {\bibinfo {author} {\bibfnamefont {J.}~\bibnamefont
  {Rembieli{\'{n}}ski}}\ and\ \bibinfo {author} {\bibfnamefont
  {P.}~\bibnamefont {Caban}},\ }\bibfield  {title} {\bibinfo {title} {Nonlinear
  extension of the quantum dynamical semigroup},\ }\href
  {https://doi.org/10.22331/q-2021-03-23-420} {\bibfield  {journal} {\bibinfo
  {journal} {{Quantum}}\ }\textbf {\bibinfo {volume} {5}},\ \bibinfo {pages}
  {420} (\bibinfo {year} {2021})}\BibitemShut {NoStop}%
\bibitem [{\citenamefont {Bargmann}(1964)}]{Bargmann}%
  \BibitemOpen
  \bibfield  {author} {\bibinfo {author} {\bibfnamefont {V.}~\bibnamefont
  {Bargmann}},\ }\bibfield  {title} {\bibinfo {title} {Note on wigner's theorem
  on symmetry operations},\ }\href {https://doi.org/10.1063/1.1704188}
  {\bibfield  {journal} {\bibinfo  {journal} {Journal of Mathematical Physics}\
  }\textbf {\bibinfo {volume} {5}},\ \bibinfo {pages} {862} (\bibinfo {year}
  {1964})}\BibitemShut {NoStop}%
\bibitem [{\citenamefont {Wigner}(1931)}]{WignerBook}%
  \BibitemOpen
  \bibfield  {author} {\bibinfo {author} {\bibfnamefont {E.~P.}\ \bibnamefont
  {Wigner}},\ }\href@noop {} {\emph {\bibinfo {title} {{Gruppentheorie und ihre
  Anwendung auf die Quanten mechanik der Atomspektren}}}}\ (\bibinfo
  {publisher} {J.W. Edwards/Friedr. Vieweg \& Sohn},\ \bibinfo {year}
  {1931})\BibitemShut {NoStop}%
\bibitem [{\citenamefont {Spekkens}(2005)}]{Spekkens05}%
  \BibitemOpen
  \bibfield  {author} {\bibinfo {author} {\bibfnamefont {R.~W.}\ \bibnamefont
  {Spekkens}},\ }\bibfield  {title} {\bibinfo {title} {Contextuality for
  preparations, transformations, and unsharp measurements},\ }\href
  {https://doi.org/10.1103/PhysRevA.71.052108} {\bibfield  {journal} {\bibinfo
  {journal} {Phys. Rev. A}\ }\textbf {\bibinfo {volume} {71}},\ \bibinfo
  {pages} {052108} (\bibinfo {year} {2005})}\BibitemShut {NoStop}%
\bibitem [{\citenamefont {Harrigan}\ and\ \citenamefont
  {Spekkens}(2010)}]{Harrigan_Spekkens}%
  \BibitemOpen
  \bibfield  {author} {\bibinfo {author} {\bibfnamefont {N.}~\bibnamefont
  {Harrigan}}\ and\ \bibinfo {author} {\bibfnamefont {R.~W.}\ \bibnamefont
  {Spekkens}},\ }\bibfield  {title} {\bibinfo {title} {Einstein,
  incompleteness, and the epistemic view of quantum states},\ }\href
  {https://doi.org/10.1007/s10701-009-9347-0} {\bibfield  {journal} {\bibinfo
  {journal} {Foundations of Physics}\ }\textbf {\bibinfo {volume} {40}},\
  \bibinfo {pages} {125} (\bibinfo {year} {2010})}\BibitemShut {NoStop}%
\bibitem [{\citenamefont {Masanes}\ \emph {et~al.}(2019)\citenamefont
  {Masanes}, \citenamefont {Galley},\ and\ \citenamefont
  {M{\"u}ller}}]{Masanes2019}%
  \BibitemOpen
  \bibfield  {author} {\bibinfo {author} {\bibfnamefont {L.}~\bibnamefont
  {Masanes}}, \bibinfo {author} {\bibfnamefont {T.~D.}\ \bibnamefont
  {Galley}},\ and\ \bibinfo {author} {\bibfnamefont {M.~P.}\ \bibnamefont
  {M{\"u}ller}},\ }\bibfield  {title} {\bibinfo {title} {The measurement
  postulates of quantum mechanics are operationally redundant},\ }\href
  {https://doi.org/10.1038/s41467-019-09348-x} {\bibfield  {journal} {\bibinfo
  {journal} {Nature Communications}\ }\textbf {\bibinfo {volume} {10}},\
  \bibinfo {pages} {1361} (\bibinfo {year} {2019})}\BibitemShut {NoStop}%
\bibitem [{\citenamefont {Stacey}(2023)}]{STACEY_critique}%
  \BibitemOpen
  \bibfield  {author} {\bibinfo {author} {\bibfnamefont {B.~C.}\ \bibnamefont
  {Stacey}},\ }\bibfield  {title} {\bibinfo {title} {{Masanes-Galley-M\"{u}ller
  and the State-Update Postulate}},\ }\href@noop {} {\  (\bibinfo {year}
  {2023})},\ \bibinfo {note} {{eprint}: \url{https://arxiv.org/abs/2211.03299 }
  [quant-ph]}\BibitemShut {NoStop}%
\bibitem [{\citenamefont {Stacey}(2022{\natexlab{a}})}]{MGM_reply}%
  \BibitemOpen
  \bibfield  {author} {\bibinfo {author} {\bibfnamefont {B.~C.}\ \bibnamefont
  {Stacey}},\ }\bibfield  {title} {\bibinfo {title} {{Masanes, Llu{\'i}s and
  Galley, Thomas D. and M{\"u}ller, Markus P.}},\ }\href@noop {} {\  (\bibinfo
  {year} {2022}{\natexlab{a}})},\ \bibinfo {note} {{eprint}:
  \url{https://arxiv.org/abs/2212.03629 } [quant-ph]}\BibitemShut {NoStop}%
\bibitem [{\citenamefont {Wigner}(1995)}]{Wigner1995}%
  \BibitemOpen
  \bibfield  {author} {\bibinfo {author} {\bibfnamefont {E.~P.}\ \bibnamefont
  {Wigner}},\ }\bibinfo {title} {Remarks on the mind-body question},\ in\ \href
  {https://doi.org/10.1007/978-3-642-78374-6_20} {\emph {\bibinfo {booktitle}
  {Philosophical Reflections and Syntheses}}},\ \bibinfo {editor} {edited by\
  \bibinfo {editor} {\bibfnamefont {J.}~\bibnamefont {Mehra}}}\ (\bibinfo
  {publisher} {Springer Berlin Heidelberg},\ \bibinfo {address} {Berlin,
  Heidelberg},\ \bibinfo {year} {1995})\ pp.\ \bibinfo {pages}
  {247--260}\BibitemShut {NoStop}%
\bibitem [{\citenamefont {Spekkens}(2019)}]{Spekkens_Leibniz}%
  \BibitemOpen
  \bibfield  {author} {\bibinfo {author} {\bibfnamefont {R.~W.}\ \bibnamefont
  {Spekkens}},\ }\bibfield  {title} {\bibinfo {title} {{The ontological
  identity of empirical indiscernibles: Leibniz's methodological principle and
  its significance in the work of Einstein}},\ }\href
  {https://arxiv.org/abs/1909.04628} {\  (\bibinfo {year} {2019})},\ \bibinfo
  {note} {eprint arXiv:1909.04628}\BibitemShut {NoStop}%
\bibitem [{\citenamefont {Healey}(2018)}]{Healey_2018}%
  \BibitemOpen
  \bibfield  {author} {\bibinfo {author} {\bibfnamefont {R.}~\bibnamefont
  {Healey}},\ }\href {http://philsci-archive.pitt.edu/14322/} {\bibinfo {title}
  {Pragmatist quantum realism}} (\bibinfo {year} {2018}),\ \bibinfo {note}
  {forthcoming in a collection on Realism and the Quantum, eds. French and
  Saatsi.}\BibitemShut {Stop}%
\bibitem [{\citenamefont {Healey}(2021)}]{Healey_2021}%
  \BibitemOpen
  \bibfield  {author} {\bibinfo {author} {\bibfnamefont {R.~A.}\ \bibnamefont
  {Healey}},\ }\href {http://philsci-archive.pitt.edu/19617/} {\bibinfo {title}
  {Representation and the quantum state}} (\bibinfo {year} {2021}),\ \bibinfo
  {note} {to appear in Valia Allori (ed.) Quantum Mechanics and Fundamentality.
  Springer.}\BibitemShut {Stop}%
\bibitem [{\citenamefont {Stacey}(2021{\natexlab{a}})}]{STACEY_RQM1}%
  \BibitemOpen
  \bibfield  {author} {\bibinfo {author} {\bibfnamefont {B.~C.}\ \bibnamefont
  {Stacey}},\ }\bibfield  {title} {\bibinfo {title} {{On Relationalist
  Reconstructions of Quantum Theory}},\ }\href@noop {} {\  (\bibinfo {year}
  {2021}{\natexlab{a}})},\ \bibinfo {note} {{eprint}:
  \url{https://arxiv.org/abs/2109.03186} [quant-ph]}\BibitemShut {NoStop}%
\bibitem [{\citenamefont {Dorato}(2020)}]{Dorato2020}%
  \BibitemOpen
  \bibfield  {author} {\bibinfo {author} {\bibfnamefont {M.}~\bibnamefont
  {Dorato}},\ }\bibfield  {title} {\bibinfo {title} {{Bohr meets Rovelli: a
  dispositionalist account of the quantum limits of knowledge}},\ }\href
  {https://doi.org/10.1007/s40509-020-00220-y} {\bibfield  {journal} {\bibinfo
  {journal} {Quantum Studies: Mathematics and Foundations}\ }\textbf {\bibinfo
  {volume} {7}},\ \bibinfo {pages} {233} (\bibinfo {year} {2020})}\BibitemShut
  {NoStop}%
\bibitem [{\citenamefont {Smerlak}\ and\ \citenamefont
  {Rovelli}(2007)}]{RQM_Smerlak}%
  \BibitemOpen
  \bibfield  {author} {\bibinfo {author} {\bibfnamefont {M.}~\bibnamefont
  {Smerlak}}\ and\ \bibinfo {author} {\bibfnamefont {C.}~\bibnamefont
  {Rovelli}},\ }\bibfield  {title} {\bibinfo {title} {Relational {EPR}},\
  }\href {https://doi.org/10.1007/s10701-007-9105-0} {\bibfield  {journal}
  {\bibinfo  {journal} {Foundations of Physics}\ }\textbf {\bibinfo {volume}
  {37}},\ \bibinfo {pages} {427} (\bibinfo {year} {2007})}\BibitemShut
  {NoStop}%
\bibitem [{\citenamefont {Di~Biagio}\ and\ \citenamefont
  {Rovelli}(2021)}]{DiBiagio_RQM}%
  \BibitemOpen
  \bibfield  {author} {\bibinfo {author} {\bibfnamefont {A.}~\bibnamefont
  {Di~Biagio}}\ and\ \bibinfo {author} {\bibfnamefont {C.}~\bibnamefont
  {Rovelli}},\ }\bibfield  {title} {\bibinfo {title} {Stable facts, relative
  facts},\ }\href {https://doi.org/10.1007/s10701-021-00429-w} {\bibfield
  {journal} {\bibinfo  {journal} {Foundations of Physics}\ }\textbf {\bibinfo
  {volume} {51}},\ \bibinfo {pages} {30} (\bibinfo {year} {2021})}\BibitemShut
  {NoStop}%
\bibitem [{\citenamefont {Adlam}\ and\ \citenamefont
  {Rovelli}(2022)}]{Adlam_RQM}%
  \BibitemOpen
  \bibfield  {author} {\bibinfo {author} {\bibfnamefont {E.}~\bibnamefont
  {Adlam}}\ and\ \bibinfo {author} {\bibfnamefont {C.}~\bibnamefont
  {Rovelli}},\ }\bibfield  {title} {\bibinfo {title} {{Information is Physical:
  Cross-Perspective Links in Relational Quantum Mechanics}},\ }\href@noop {} {\
   (\bibinfo {year} {2022})},\ \bibinfo {note} {{eprint}:
  \url{https://arxiv.org/abs/203.13342} [quant-ph]}\BibitemShut {NoStop}%
\bibitem [{\citenamefont {Stacey}(2021{\natexlab{b}})}]{STACEY_RQM2}%
  \BibitemOpen
  \bibfield  {author} {\bibinfo {author} {\bibfnamefont {B.~C.}\ \bibnamefont
  {Stacey}},\ }\bibfield  {title} {\bibinfo {title} {{Is Relational Quantum
  Mechanics about Facts? If So, Whose? A Reply to Di Biagio and Rovelli's
  Comment on Brukner and Pienaar}},\ }\href@noop {} {\  (\bibinfo {year}
  {2021}{\natexlab{b}})},\ \bibinfo {note} {{eprint}:
  \url{https://arxiv.org/abs/2112.07830} [quant-ph]}\BibitemShut {NoStop}%
\bibitem [{\citenamefont {Stacey}(2022{\natexlab{b}})}]{STACEY_RQM3}%
  \BibitemOpen
  \bibfield  {author} {\bibinfo {author} {\bibfnamefont {B.~C.}\ \bibnamefont
  {Stacey}},\ }\bibfield  {title} {\bibinfo {title} {{The De-Relationalizing of
  Relational Quantum Mechanics}},\ }\href@noop {} {\  (\bibinfo {year}
  {2022}{\natexlab{b}})},\ \bibinfo {note} {{eprint}:
  \url{https://arxiv.org/abs/2211.03230} [quant-ph]}\BibitemShut {NoStop}%
\bibitem [{\citenamefont {Rovelli}(1996)}]{ROVELLI_96}%
  \BibitemOpen
  \bibfield  {author} {\bibinfo {author} {\bibfnamefont {C.}~\bibnamefont
  {Rovelli}},\ }\bibfield  {title} {\bibinfo {title} {{Relational Quantum
  Mechanics}},\ }\href {https://doi.org/10.1007/BF02302261} {\bibfield
  {journal} {\bibinfo  {journal} {International Journal of Theoretical
  Physics}\ }\textbf {\bibinfo {volume} {35}},\ \bibinfo {pages} {1637}
  (\bibinfo {year} {1996})}\BibitemShut {NoStop}%
\bibitem [{\citenamefont {Laudisa}\ and\ \citenamefont
  {Rovelli}(2021)}]{SEP_RQM}%
  \BibitemOpen
  \bibfield  {author} {\bibinfo {author} {\bibfnamefont {F.}~\bibnamefont
  {Laudisa}}\ and\ \bibinfo {author} {\bibfnamefont {C.}~\bibnamefont
  {Rovelli}},\ }\bibfield  {title} {\bibinfo {title} {{Relational Quantum
  Mechanics}},\ }in\ \href@noop {} {\emph {\bibinfo {booktitle} {The {Stanford}
  Encyclopedia of Philosophy}}},\ \bibinfo {editor} {edited by\ \bibinfo
  {editor} {\bibfnamefont {E.~N.}\ \bibnamefont {Zalta}}}\ (\bibinfo
  {publisher} {Metaphysics Research Lab, Stanford University},\ \bibinfo {year}
  {2021})\ \bibinfo {edition} {spring 2021}\ ed.\BibitemShut {Stop}%
\bibitem [{\citenamefont {Auff{\`e}ves}\ and\ \citenamefont
  {Grangier}(2020)}]{Auffeves2020}%
  \BibitemOpen
  \bibfield  {author} {\bibinfo {author} {\bibfnamefont {A.}~\bibnamefont
  {Auff{\`e}ves}}\ and\ \bibinfo {author} {\bibfnamefont {P.}~\bibnamefont
  {Grangier}},\ }\bibfield  {title} {\bibinfo {title} {Deriving born's rule
  from an inference to the best explanation},\ }\href
  {https://doi.org/10.1007/s10701-020-00326-8} {\bibfield  {journal} {\bibinfo
  {journal} {Foundations of Physics}\ }\textbf {\bibinfo {volume} {50}},\
  \bibinfo {pages} {1781} (\bibinfo {year} {2020})}\BibitemShut {NoStop}%
\bibitem [{\citenamefont {Auff{\`e}ves}\ and\ \citenamefont
  {Grangier}(2022)}]{Auffeves_2022}%
  \BibitemOpen
  \bibfield  {author} {\bibinfo {author} {\bibfnamefont {A.}~\bibnamefont
  {Auff{\`e}ves}}\ and\ \bibinfo {author} {\bibfnamefont {P.}~\bibnamefont
  {Grangier}},\ }\bibfield  {title} {\bibinfo {title} {Revisiting born's rule
  through uhlhorn's and gleason's theorems},\ }\bibfield  {journal} {\bibinfo
  {journal} {Entropy}\ }\textbf {\bibinfo {volume} {24}},\ \href
  {https://doi.org/10.3390/e24020199} {10.3390/e24020199} (\bibinfo {year}
  {2022})\BibitemShut {NoStop}%
\bibitem [{\citenamefont {Leifer}(2014)}]{Leifer_review}%
  \BibitemOpen
  \bibfield  {author} {\bibinfo {author} {\bibfnamefont {M.}~\bibnamefont
  {Leifer}},\ }\bibfield  {title} {\bibinfo {title} {Is the quantum state real?
  an extended review of psi-ontology theorems},\ }\href
  {https://doi.org/10.12743/quanta.v3i1.22} {\bibfield  {journal} {\bibinfo
  {journal} {Quanta}\ }\textbf {\bibinfo {volume} {3}},\ \bibinfo {pages} {67}
  (\bibinfo {year} {2014})}\BibitemShut {NoStop}%
\bibitem [{\citenamefont {Bub}\ and\ \citenamefont
  {Pitowsky}(2010)}]{BUB_PITOWSKI}%
  \BibitemOpen
  \bibfield  {author} {\bibinfo {author} {\bibfnamefont {J.}~\bibnamefont
  {Bub}}\ and\ \bibinfo {author} {\bibfnamefont {I.}~\bibnamefont {Pitowsky}},\
  }\bibfield  {title} {\bibinfo {title} {Two dogmas about quantum mechanics},\
  }in\ \href {https://arxiv.org/abs/0712.4258} {\emph {\bibinfo {booktitle}
  {{Many Worlds?: Everett, Quantum Theory, \& Reality}}}},\ \bibinfo {editor}
  {edited by\ \bibinfo {editor} {\bibfnamefont {S.}~\bibnamefont {Saunders}},
  \bibinfo {editor} {\bibfnamefont {J.}~\bibnamefont {Barrett}}, \bibinfo
  {editor} {\bibfnamefont {A.}~\bibnamefont {Kent}},\ and\ \bibinfo {editor}
  {\bibfnamefont {D.}~\bibnamefont {Wallace}}}\ (\bibinfo  {publisher} {{Oxford
  University Press}},\ \bibinfo {year} {2010})\BibitemShut {NoStop}%
\bibitem [{\citenamefont {Bub}(2019)}]{BUB_19}%
  \BibitemOpen
  \bibfield  {author} {\bibinfo {author} {\bibfnamefont {J.}~\bibnamefont
  {Bub}},\ }\bibfield  {title} {\bibinfo {title} {{Two dogmas' redux}},\
  }\href@noop {} {\  (\bibinfo {year} {2019})},\ \bibinfo {note} {{eprint}:
  \url{https://arxiv.org/abs/1907.06240} [quant-ph]}\BibitemShut {NoStop}%
\bibitem [{\citenamefont {Fuchs}(2010)}]{Fuchs10a}%
  \BibitemOpen
  \bibfield  {author} {\bibinfo {author} {\bibfnamefont {C.~A.}\ \bibnamefont
  {Fuchs}},\ }\bibfield  {title} {\bibinfo {title} {{QBism, the Perimeter of
  Quantum Bayesianism}},\ }\href@noop {} {\  (\bibinfo {year} {2010})},\
  \bibinfo {note} {{eprint}: \url{https://arxiv.org/abs/1003.5209}
  [quant-ph]}\BibitemShut {NoStop}%
\bibitem [{\citenamefont {Fuchs}\ \emph {et~al.}(2014)\citenamefont {Fuchs},
  \citenamefont {Mermin},\ and\ \citenamefont {Schack}}]{QBism_FDR2014}%
  \BibitemOpen
  \bibfield  {author} {\bibinfo {author} {\bibfnamefont {C.~A.}\ \bibnamefont
  {Fuchs}}, \bibinfo {author} {\bibfnamefont {N.~D.}\ \bibnamefont {Mermin}},\
  and\ \bibinfo {author} {\bibfnamefont {R.}~\bibnamefont {Schack}},\
  }\bibfield  {title} {\bibinfo {title} {An introduction to qbism with an
  application to the locality of quantum mechanics},\ }\href
  {https://doi.org/10.1119/1.4874855} {\bibfield  {journal} {\bibinfo
  {journal} {American Journal of Physics}\ }\textbf {\bibinfo {volume} {82}},\
  \bibinfo {pages} {749} (\bibinfo {year} {2014})},\ \Eprint
  {https://arxiv.org/abs/https://doi.org/10.1119/1.4874855}
  {https://doi.org/10.1119/1.4874855} \BibitemShut {NoStop}%
\bibitem [{\citenamefont {Fuchs}\ and\ \citenamefont
  {Stacey}(2018)}]{FuchsStacey2018}%
  \BibitemOpen
  \bibfield  {author} {\bibinfo {author} {\bibfnamefont {C.~A.}\ \bibnamefont
  {Fuchs}}\ and\ \bibinfo {author} {\bibfnamefont {B.~C.}\ \bibnamefont
  {Stacey}},\ }\bibfield  {title} {\bibinfo {title} {{QBism: Quantum Theory as
  a Hero's Handbook}},\ }in\ \href@noop {} {\emph {\bibinfo {booktitle}
  {{Proceedings of the International School of Physics ``Enrico Fermi'' Course
  197 -- Foundations of Quantum Physics}}}},\ \bibinfo {editor} {edited by\
  \bibinfo {editor} {\bibfnamefont {E.~M.}\ \bibnamefont {Rasel}}, \bibinfo
  {editor} {\bibfnamefont {W.~P.}\ \bibnamefont {Schleich}},\ and\ \bibinfo
  {editor} {\bibfnamefont {S.}~\bibnamefont {W\"{o}lk}}}\ (\bibinfo
  {publisher} {IOS Press, Amsterdam; Societ\`a Italiana di Fisica, Bologna,
  2018},\ \bibinfo {year} {2018})\ pp.\ \bibinfo {pages} {133--202}\BibitemShut
  {NoStop}%
\bibitem [{\citenamefont {Bong}\ \emph {et~al.}(2020)\citenamefont {Bong},
  \citenamefont {Utreras-Alarc{\'o}n}, \citenamefont {Ghafari}, \citenamefont
  {Liang}, \citenamefont {Tischler}, \citenamefont {Cavalcanti}, \citenamefont
  {Pryde},\ and\ \citenamefont {Wiseman}}]{BONG_2020}%
  \BibitemOpen
  \bibfield  {author} {\bibinfo {author} {\bibfnamefont {K.-W.}\ \bibnamefont
  {Bong}}, \bibinfo {author} {\bibfnamefont {A.}~\bibnamefont
  {Utreras-Alarc{\'o}n}}, \bibinfo {author} {\bibfnamefont {F.}~\bibnamefont
  {Ghafari}}, \bibinfo {author} {\bibfnamefont {Y.-C.}\ \bibnamefont {Liang}},
  \bibinfo {author} {\bibfnamefont {N.}~\bibnamefont {Tischler}}, \bibinfo
  {author} {\bibfnamefont {E.~G.}\ \bibnamefont {Cavalcanti}}, \bibinfo
  {author} {\bibfnamefont {G.~J.}\ \bibnamefont {Pryde}},\ and\ \bibinfo
  {author} {\bibfnamefont {H.~M.}\ \bibnamefont {Wiseman}},\ }\bibfield
  {title} {\bibinfo {title} {A strong no-go theorem on the {W}igner's friend
  paradox},\ }\href {https://doi.org/10.1038/s41567-020-0990-x} {\bibfield
  {journal} {\bibinfo  {journal} {Nature Physics}\ }\textbf {\bibinfo {volume}
  {16}},\ \bibinfo {pages} {1199} (\bibinfo {year} {2020})}\BibitemShut
  {NoStop}%
\bibitem [{\citenamefont {Cavalcanti}(2021)}]{Cavalcanti_Bubble}%
  \BibitemOpen
  \bibfield  {author} {\bibinfo {author} {\bibfnamefont {E.~G.}\ \bibnamefont
  {Cavalcanti}},\ }\bibfield  {title} {\bibinfo {title} {The view from a wigner
  bubble},\ }\href {https://doi.org/10.1007/s10701-021-00417-0} {\bibfield
  {journal} {\bibinfo  {journal} {Foundations of Physics}\ }\textbf {\bibinfo
  {volume} {51}},\ \bibinfo {pages} {39} (\bibinfo {year} {2021})}\BibitemShut
  {NoStop}%
\bibitem [{\citenamefont {Cavalcanti}\ and\ \citenamefont
  {Wiseman}(2021)}]{Cavalcanti_Qcausality_2021}%
  \BibitemOpen
  \bibfield  {author} {\bibinfo {author} {\bibfnamefont {E.~G.}\ \bibnamefont
  {Cavalcanti}}\ and\ \bibinfo {author} {\bibfnamefont {H.~M.}\ \bibnamefont
  {Wiseman}},\ }\bibfield  {title} {\bibinfo {title} {Implications of local
  friendliness violation for quantum causality},\ }\bibfield  {journal}
  {\bibinfo  {journal} {Entropy}\ }\textbf {\bibinfo {volume} {23}},\ \href
  {https://doi.org/10.3390/e23080925} {10.3390/e23080925} (\bibinfo {year}
  {2021})\BibitemShut {NoStop}%
\bibitem [{\citenamefont {Spekkens}(2015)}]{spekkens_paradigm_2015}%
  \BibitemOpen
  \bibfield  {author} {\bibinfo {author} {\bibfnamefont {R.~W.}\ \bibnamefont
  {Spekkens}},\ }\bibfield  {title} {\bibinfo {title} {The {Paradigm} of
  {Kinematics} and {Dynamics} {Must} {Yield} to {Causal} {Structure}},\ }in\
  \href {https://doi.org/10.1007/978-3-319-13045-3_2} {\emph {\bibinfo
  {booktitle} {Questioning the {Foundations} of {Physics}: {Which} of {Our}
  {Fundamental} {Assumptions} {Are} {Wrong}?}}},\ \bibinfo {series and number}
  {The {Frontiers} {Collection}},\ \bibinfo {editor} {edited by\ \bibinfo
  {editor} {\bibfnamefont {A.}~\bibnamefont {Aguirre}}, \bibinfo {editor}
  {\bibfnamefont {B.}~\bibnamefont {Foster}},\ and\ \bibinfo {editor}
  {\bibfnamefont {Z.}~\bibnamefont {Merali}}}\ (\bibinfo  {publisher} {Springer
  International Publishing},\ \bibinfo {address} {Cham},\ \bibinfo {year}
  {2015})\ pp.\ \bibinfo {pages} {5--16}\BibitemShut {NoStop}%
\bibitem [{\citenamefont {Ghirardi}\ \emph {et~al.}(1986)\citenamefont
  {Ghirardi}, \citenamefont {Rimini},\ and\ \citenamefont {Weber}}]{GRW}%
  \BibitemOpen
  \bibfield  {author} {\bibinfo {author} {\bibfnamefont {G.~C.}\ \bibnamefont
  {Ghirardi}}, \bibinfo {author} {\bibfnamefont {A.}~\bibnamefont {Rimini}},\
  and\ \bibinfo {author} {\bibfnamefont {T.}~\bibnamefont {Weber}},\ }\bibfield
   {title} {\bibinfo {title} {Unified dynamics for microscopic and macroscopic
  systems},\ }\href {https://doi.org/10.1103/PhysRevD.34.470} {\bibfield
  {journal} {\bibinfo  {journal} {Phys. Rev. D}\ }\textbf {\bibinfo {volume}
  {34}},\ \bibinfo {pages} {470} (\bibinfo {year} {1986})}\BibitemShut
  {NoStop}%
\bibitem [{\citenamefont {Pearle}(1989)}]{Pearle_CSL}%
  \BibitemOpen
  \bibfield  {author} {\bibinfo {author} {\bibfnamefont {P.}~\bibnamefont
  {Pearle}},\ }\bibfield  {title} {\bibinfo {title} {Combining stochastic
  dynamical state-vector reduction with spontaneous localization},\ }\href
  {https://doi.org/10.1103/PhysRevA.39.2277} {\bibfield  {journal} {\bibinfo
  {journal} {Phys. Rev. A}\ }\textbf {\bibinfo {volume} {39}},\ \bibinfo
  {pages} {2277} (\bibinfo {year} {1989})}\BibitemShut {NoStop}%
\bibitem [{\citenamefont {Jordan}(1993)}]{JORDAN_1993}%
  \BibitemOpen
  \bibfield  {author} {\bibinfo {author} {\bibfnamefont {T.}~\bibnamefont
  {Jordan}},\ }\bibfield  {title} {\bibinfo {title} {{Reconstructing a
  Nonlinear Dynamical Framework for Testing Quantum Mechanics}},\ }\href
  {https://doi.org/https://doi.org/10.1006/aphy.1993.1053} {\bibfield
  {journal} {\bibinfo  {journal} {Annals of Physics}\ }\textbf {\bibinfo
  {volume} {225}},\ \bibinfo {pages} {83} (\bibinfo {year} {1993})}\BibitemShut
  {NoStop}%
\bibitem [{\citenamefont {Haag}\ and\ \citenamefont
  {Bannier}(1978)}]{Haag1978}%
  \BibitemOpen
  \bibfield  {author} {\bibinfo {author} {\bibfnamefont {R.}~\bibnamefont
  {Haag}}\ and\ \bibinfo {author} {\bibfnamefont {U.}~\bibnamefont {Bannier}},\
  }\bibfield  {title} {\bibinfo {title} {{Comments on Mielnik's generalized
  (non linear) quantum mechanics}},\ }\href
  {https://doi.org/10.1007/BF01609470} {\bibfield  {journal} {\bibinfo
  {journal} {Communications in Mathematical Physics}\ }\textbf {\bibinfo
  {volume} {60}},\ \bibinfo {pages} {1} (\bibinfo {year} {1978})}\BibitemShut
  {NoStop}%
\end{thebibliography}
\end{document}